\documentclass[aps,prb,reprint,groupedaddress]{revtex4-2}
\usepackage{amsmath}
\usepackage{amssymb}
\usepackage{graphicx}
\usepackage{subfigure}
\usepackage{comment}
\usepackage{ntheorem}
\begin{document}
\newtheorem{Definition}{Definition}[subsection]
   \title{Topological Classification of Gapped/Gap-preserving Rational Space-Time Crystal Systems}
   \author{Shu-Xuan Wang }
   \email{wshx123@mail.ustc.edu.cn}
   \affiliation{Department of Modern Physics, University of Science and Technology of China, Hefei, 230026, China}
   \author{Shaolong Wan}
   \email{slwan@ustc.edu.cn}
    \affiliation{Department of Modern Physics, University of Science and Technology of China, Hefei, 230026, China}
    
   \date{\today}
   
   \begin{abstract}
    The traditional systems researched in condensed matter physics always have spatial translation symmetry. However for space-time crystal systems, the spatial translation symmetry is no longer preserved and the lattice potential have space-time translation symmetry instead. We show that a rational space-time crystal system is equal to a traditional floquet system. Then, we find a way to solve the floquet equation analytically and construct an effective Hamiltonian of the rational space-time crystal system. By this effective Hamiltonian, we obtain the topological classification of gapped and gap-preserving rational space-time crystal systems. Our works reveal the correlation between space-time crystal systems and floquet systems and give a systematic method to explore the properties of rational space-time crystal systems. 
   \end{abstract}
   
   \maketitle

   \section{Introduction}
      Topological phase is an important topic in condensed matter physics~\cite{10,11,12,13,14,15,16,17,18,19,20,21,22,23,25,26,27,4,5,6,7,8}. There are numerous novel phenomena about topological phase. For a topological non-trivial system, there exist some states localized at boundaries, whose energies are isolated in the spectrum, which are called topological states~\cite{11,26}. The topological states can apply conductance even if the bulk of the system is gapped, and this phenomenon can be described by an effective field theory, topological field theory~\cite{11,12}. If the system has some crystalline symmetries, the topological states may no longer be localized at the boundary, but localized at the corner, these topological states are called high-order topological states~\cite{16,17,18}. Theoretically, for a given system, we can use the topological invariants to describe whether this system is topological non-trivial or not~\cite{11,12,26}, and a topological non-trivial system can not be transformed to a topological trivial system adiabatically without an energy gap closing. Another fascinating direction is classification of the topological phase. For systems with different internal or crystalline symmetries, the topological invariants used to describe these topological system may be different. For traditional Hermitian and non-Hermitian topological systems, the classification and topological invariants can be given by $K$ theory according to the symmetries and the type of the energy gap of the systems~\cite{10,8}.
      \par

      In the past, the topological systems researched always have the spatial translation symmetry. Recently, space-time crystal system has been noticed~\cite{1,2}. Spatial translation symmetry is no longer preserved in space-time crystal systems. Instead, space-time translation symmetry is kept in these systems. Hence, Bloch theorem is invalid and  generalized floquet-Bloch theorem is used in space-time crystal systems ~\cite{1}. Similar to traditional systems, it is natural to expect that space-time crystal systems can have non-trivial topological phase, which can be described by topological invariants.
      \par 

      In this article, we show that a rational space-time crystal system can be transformed to a traditional floquet system, which respect the spatial transition symmetry.  Solving the floquet equation, we can obtain the quasi-spectrum of the rational space-time crystal system and construct an effective Hamiltonian. Then, topological classification of rational space-time crystal systems is reduced to the topological classification of the effective Hamiltonian. Thus, by researching the effective Hamiltonian, we obtain the topological classification of gapped/gap-preserving rational space-time crystal systems.
      \par

      This article is organized as follows. In Sec.$\mathbf{II}$, we give the definition of rational space-time crystal system and show that a rational space-time crystal system is equal to a traditional floquet system. In Sec.$\mathbf{III}$, we solve the floquet equation and construct the effective Hamiltonian. In Sec.$\mathbf{IV}$, we give the transformation of the effective Hamiltonian under internal symmetries. In Sec.$\mathbf{V}$, we discuss the type of energy gap about Hermitian and non-Hermitian rational space-time crystal systems and give the topological classification of gapped/gap-preserving rational space-time crystal systems. In Sec.$\mathbf{VI}$, we give examples of topological classification about Hermitian and non-Hermitian space-time crystal system. Conclusion and discussion is given in Sec.$\mathbf{VII}$.

   \section{Space-Time translation symmetry and Hamiltonian of Rational Space-Time Crystal}
      For $(d+1)$-dimensional space-time crystal, the lattice potential may not satisfy the translation symmetry. Instead, it satisfies space-time translation symmetry~\cite{1,2}:
        \begin{equation}
            V(\mathbf{r},t) = V(\mathbf{r} + \mathbf{s}_i, t + \tau_i)
            \label{1}
        \end{equation} 
      for $i=0,1, \cdots, d$. Since $\mathbf{s}_i$ is a $d$-dimensional spatial vector, only $d$ vectors of $\{ \mathbf{s}_i \}$ are independent. Hence, there exist $m_i \in \mathbb{R}$ for $i=0,1,2, \cdots, d$, such that\footnote[1]{at least one number of $\{ m_i \}$ is nonzero.}
        \begin{equation}
           \sum_{i=0}^d m_i \mathbf{s}_i = 0.
            \label{2}
        \end{equation} 
      We call the space-time crystal system is rational space-time crystal system if $m_i \in \mathbb{Z}$ for $i=0,1,2,\cdots,d$ and $\frac{\tau_i}{\tau_j}$ is rational number for all $i,j = 0,1,2,\cdots,d$. For the rational space-time crystal system, we can
      define $T =  \sum_{i=0}^d m_i \tau_i$. Then we have
        \begin{equation}
            V(\mathbf{r},t) = V(\mathbf{r} + \sum_{i=0}^d m_i \mathbf{s}_i , t + \sum_{i=0}^d m_i \tau_i)  = V(\mathbf{r} , t + T).
            \label{3}
        \end{equation}
    It means the potential of the rational space-time crystal system has a period, $T$, about time. Thus, the discrete space-time translation basis vectors can be written as 
       \begin{equation}
         \mathbf{a}_0 = (\mathbf{0}, -T), \qquad \mathbf{a}_i = (\mathbf{s}_i, \alpha_i T),
         \label{4}
       \end{equation}
    with $\alpha_i = \frac{\tau_i}{T} \ \mathrm{mod} \ 1 = (\sum_{j=0}^d m_j \frac{\tau_j}{\tau_i})^{-1} \ \mathrm{mod} \ 1 \in (-\frac{1}{2},\frac{1}{2}]$ for $i=1,2, \cdots, d$. Since $m_j$ is integer and $\frac{\tau_j}{\tau_i}$ is rational for all $i=1,2,\cdots,d$ and $j=0,1,2,\cdots,d$, $\alpha_i$ is rational number for $i=1,2,\cdots,d$. The reciprocal vectors corresponding to the discrete space-time translation basis vectors are 
       \begin{equation}
        \mathbf{b}_0 = (\sum_{i=1}^d \alpha_i \mathbf{g}_i, \Omega),  \qquad
        \mathbf{b}_i = (\mathbf{g}_i, 0),
        \label{5}
       \end{equation}
    with $\Omega = \frac{2 \pi}{T}$, $\mathbf{g}_i \cdot \mathbf{s}_j = 2\pi \delta_{i,j}$. Hence, $\mathbf{a}_i \cdot \mathbf{b}_j = 2\pi \delta_{i,j}$ \footnote[2]{We take $\mathbf{g}= diag \{ \mathbf{1}_{d \times d} , -1 \}$ as the space-time metric to make it consistent with Refs.\cite{1},\cite{2} }. Then, we take $\{ \mathbf{s}_i \}$ as lattice vectors, and the tight-binding Hamiltonian of the system is~\cite{2}
      \begin{gather}
        H = \sum_{n=- \infty}^{\infty} e^{- i n \Omega t} H_n,
        \label{6}  \\
        H_n = \sum_{\mathbf{R}_i,\mathbf{R}_j} e^{i n  \mathbf{ \delta k} \cdot \mathbf{R}_i} t_{\mathbf{R}_i-\mathbf{R}_j}^{(n)} c_{\mathbf{R}_i}^{\dagger} c_{\mathbf{R}_j} ,
        \label{7}
      \end{gather}
    where $\mathbf{\delta k} = \sum_{i=1}^d \alpha_i \mathbf{g}_i $, $c_{\mathbf{R}_i}$ and $c_{\mathbf{R}_i}^{\dagger}$ are creation and annihilation operator about the site locating at $\mathbf{R}_i$, and $t_{\mathbf{R}_i-\mathbf{R}_j}^{(n)}$ is the hopping amplitude between site $\mathbf{R}_i$ and $\mathbf{R}_j$, which only depends on the value of $\mathbf{R}_i-\mathbf{R}_j$.
    \par

    Since $\alpha_i$ is rational number, we can take $\alpha_i = \frac{p_i}{q_i}$, where $p_i$ and $q_i$ are coprime for $i = 1, 2, \cdots, d$. In this case, we can treat $\prod_{i=1}^d q_i$ sites as $\prod_{i=1}^d q_i$ orbitals in a unit cell. Hence, the Hamiltonian can be rewritten as~(See Appendix A for details):
       \begin{equation}
          \begin{split}
            H  & = \sum_{n=- \infty}^{\infty} e^{- i n \Omega t} \sum_{\tilde{\mathbf{R}}_i, \tilde{\mathbf{R}}_j} \mathbf{c}_{\tilde{\mathbf{R}}_i}^{\dagger} \mathbf{M}_{\tilde{\mathbf{R}}_i - \tilde{\mathbf{R}}_j}^{(n)} \mathbf{c}_{\tilde{\mathbf{R}}_j}     \\
            & = \sum_{n=- \infty}^{\infty} e^{- i n \Omega t} \tilde{H}_n 
          \end{split} 
        ,
         \label{8}
       \end{equation} 
    where $\mathbf{M}_{i-j}^{(n)}$ is a $(\prod_{i=1}^d q_i) \times (\prod_{i=1}^d q_i$) matrix only dependent on $\tilde{\mathbf{R}}_i - \tilde{\mathbf{R}}_j $ and $\mathbf{c}$~($\mathbf{c}^{\dagger}$) is a $(\prod_{i=1}^d q_i)$-dimensional column vector composed by annihilation~(creation) operators. 
    \par 

    The spatial translation symmetry is respected in Eq.\eqref{8}. Hence, a rational space-time crystal system is equal to a floquet system.

  \section{Qusi-spectrum and effect Hamiltonian of Rational Space-time crystal system}
    We start from the Hamiltonian given in Eq.\eqref{8}. The Schr\"{o}dinger equation about the Hamiltonian, $H$, is 
       \begin{equation}
          H | \psi (t) \rangle = i \partial_t | \psi (t) \rangle .
          \label{9}
       \end{equation}
    Since $H$ has a period $T$, the solution of Eq.\eqref{9} has the form\cite{3,4,24}
       \begin{equation}
        | \psi (t) \rangle = e^{-i \omega t} | u_{\omega} (t) \rangle ,
        \label{10}
       \end{equation}
    where $\omega$ is the quasienergy and $| u_{\omega} (t) \rangle$ is quasistate with period $T$. Since
       \begin{equation}
        | u_{\omega} (t) \rangle = | u_{\omega} (t+T) \rangle ,
        \label{11}
       \end{equation}
    $| u_{\omega} (t) \rangle$ can be decomposed as 
       \begin{equation}
        | u_{\omega} (t) \rangle = \sum_{m= - \infty}^{\infty} e^{- i m \Omega t} | u_{m,\omega}  \rangle  ,
        \label{12}
       \end{equation}
    with $\Omega = \frac{2\pi}{T}$. Substituting Eqs.\eqref{8},\eqref{10},\eqref{12} into Eq.\eqref{9}, we obtain 
       \begin{multline}
        \sum_{m,n} (\tilde{H}_n e^{- i (m+n) \Omega t} - \delta_{m,n} e^{- i m \Omega t} m \Omega) | u_{m,\omega} \rangle   \\ 
        = \omega \sum_m e^{- i m \Omega t} | u_{m,\omega} \rangle .
        \label{13}
       \end{multline}
    Then, we can get the equation about $| u_{m,\omega} \rangle$,
       \begin{equation}
        \sum_{n=-\infty}^{\infty} \tilde{H}_{m-n} | u_{n,\omega} \rangle - m \Omega | u_{m,\omega} \rangle = \omega | u_{m,\omega} \rangle.
        \label{14}
       \end{equation}
    The eigenvalue problem about Eq.\eqref{14} has a property, if $\omega$ is the quasienergy of $| u_{m,\omega} \rangle$, $\omega + n \Omega$ is the quasienergy of $| u_{m+n,\omega} \rangle$ \cite{5}. We call the set of quasienergies of $| u_{m,\omega} \rangle$ the $m$th sector of the quasi-spectrum.
    \par

    We define an enlarged Hamiltonian,
       \begin{equation}
        H_{enlarged} = 
           \begin{pmatrix}
             \ddots  &    &     &   & \\
                     &  \tilde{H}_0 - \Omega & \tilde{H}_{-1} & \tilde{H}_{-2} & \\
                     &  \tilde{H}_{1} & \tilde{H}_{0} & \tilde{H}_{-1} &  \\
                     & \tilde{H}_{2} & \tilde{H}_{1} & \tilde{H}_{0} + \Omega & \\
                     &   &   &  & \ddots
           \end{pmatrix},
           \label{15}
       \end{equation}
    and an enlarged vector,
       \begin{equation}
          \mathbf{u}_{enlarged}^T (\omega) = (\cdots, | u_{-1,\omega} \rangle^T, | u_{0,\omega} \rangle^T, | u_{1,\omega} \rangle^T, \cdots ).
          \label{16}
       \end{equation}
    Then, Eq,\eqref{14} can be written as
       \begin{equation}
         H_{enlarged} \mathbf{u}_{enlarged} (\omega) = \omega \mathbf{u}_{enlarged} (\omega) ,
         \label{17}
       \end{equation}
    and the set of eigenvalues of Eq.\eqref{17} is the $0$th sector of the quasi-spectrum. Since the index of sector, $m$, varies from $-\infty$ to $\infty$ in Eq.\eqref{16}, the boundary condition of $H_{enlarged}$ about sector is infinite boundary condition~(IBC) and $H_{enlarged}$ is an infinite-dimensional matrix. Thus, we assume that
       \begin{equation}
          | u_{m,\omega} \rangle = e^{- i m \theta} | u_{0,\omega} \rangle ,
          \label{18}
       \end{equation}
    where $\theta \in [0,2\pi]$. Then, Eq.\eqref{17} can be reduced as 
       \begin{equation}
          \sum_{n= - \infty}^{\infty} \tilde{H}_{m-n} e^{- i (n-m) \theta} | u_{m,\omega} \rangle = (\omega + m \Omega) | u_{m,\omega} \rangle.
          \label{19}
       \end{equation}
    For the $0$th sector,
       \begin{equation}
          \sum_{n= - \infty}^{\infty} \tilde{H}_{-n} e^{- i n \theta} | u_{0,\omega} \rangle = \omega | u_{0,\omega} \rangle.
          \label{20}
       \end{equation}
    Hence, we define an effective Hamiltonian
       \begin{equation}
          H_{eff} (\theta)= \sum_{n= - \infty}^{\infty} \tilde{H}_{-n} e^{- i n \theta}.
          \label{21}
       \end{equation}
    With $\theta$ varying from $0$ to $2 \pi$, all eigenvalues of $H_{eff} (\theta)$ compose the $0$th sector of the quasi-spectrum. In this work, we only consider the $0$th sector, because $0$th sector gives the physical spectrum~(Appendix B).

   \section{Internal Symmetry of the Effective Hamiltonian }

       Now, we discuss the internal symmetries of the system. We use $\mathcal{T}$, $\mathcal{C}$ and $\mathcal{S}$ to denote the time-reversal operator, particle-hole operator and  chiral operator. Here, we consider fermion systems. If the system has time-reversal symmetry~(TRS), the effective Hamiltonian has the relation
         \begin{equation}
           \mathcal{T} H_{eff} (\theta) \mathcal{T}^{-1} = H_{eff} (\theta) ,
           \label{22}
         \end{equation}
       Under periodic boundary condition~(PBC) about the spatial dimension, we can transform the Hamiltonian to the momentum space, and the relation Eq.\eqref{22} becomes
         \begin{equation}
            \mathcal{T} H_{eff} (\mathbf{k},\theta) \mathcal{T}^{-1} = H_{eff} (-\mathbf{k},\theta),
            \label{23}
         \end{equation} 
       where
         \begin{equation}
            H_{eff} (\mathbf{k},\theta) = \sum_{n= - \infty}^{\infty} \tilde{H}_{-n}(\mathbf{k}) e^{- i n \theta} .
            \label{24}
         \end{equation} 
       For particle-hole symmetry~(PHS) and chiral symmetry~(CS), the relations are~(See Appendix C for details) 
         \begin{gather}
            \mathcal{C} H_{eff} (\mathbf{k},\theta) \mathcal{C}^{-1} =- H_{eff} (-\mathbf{k},\theta)  
            \label{25}  \\
            \mathcal{S} H_{eff} (\mathbf{k},\theta) \mathcal{S}^{-1} = -H_{eff} (\mathbf{k},\theta).
            \label{26}
         \end{gather}
       \par

       These relations can be generalized to the internal symmetries of non-Hermitian cases~\cite{8}~(Appendix C). These relations are
         \begin{gather}
            \mathcal{T}_{+} H_{eff} (\mathbf{k},\theta) \mathcal{T}^{-1}_{+} = H_{eff} (-\mathbf{k},\theta)   \label{27}   \\
            \mathcal{C}_{-} H_{eff} (\mathbf{k},\theta) \mathcal{C}^{-1}_{-} = -H_{eff} (-\mathbf{k},\theta)    \label{28}  \\
            \mathcal{C}_{+} H_{eff}^{\dagger} (\mathbf{k},\theta) \mathcal{C}^{-1}_{+} = H_{eff} (-\mathbf{k},\theta)   \label{29}   \\
            \mathcal{T}_{-} H_{eff}^{\dagger} (\mathbf{k},\theta) \mathcal{T}^{-1}_{-} = -H_{eff} (-\mathbf{k},\theta)   \label{30}   \\
            \mathcal{S} H_{eff} (\mathbf{k},\theta) \mathcal{S}^{-1} = -H_{eff} (\mathbf{k},\theta)    \label{31}   \\
            \Gamma H_{eff}^{\dagger} (\mathbf{k},\theta) \Gamma^{-1} = - H_{eff} (\mathbf{k},\theta)        \label{32}  \\
            \eta H_{eff}^{\dagger} (\mathbf{k},\theta) \eta^{-1} = H_{eff} (\mathbf{k},\theta)
            \label{33 }  .
         \end{gather}
       Here, $\mathcal{T}_{+}$ and $\mathcal{C}_{+}$ are operators of TRS and TRS$^{\dagger}$, $\mathcal{C}_{-}$ and $\mathcal{T}_{-}$ are operators of PHS and PHS$^{\dagger}$, $\mathcal{S}$ is the operator of CS, $\Gamma$ is the operator of sublattice symmetry~(SLS), and $\eta$ is the operator of pseudo-Hermitian symmetry.
       \par

       Under internal symmetry transformations, $\theta$ is always invariant. Hence, we treat $\theta$ as a parameter but not a momentum-like variable even if $\theta$ has a period, $2 \pi$.
       \par

       In addition, it is worth to point out that TRS is not independent of the space-time translation symmetry. TRS gives constraints about space-time translation symmetry. A breif discussion about these constraints is given in Appendix D.

   \section{Topological Classification of According to the Effective Hamiltonian}
      Since $H_{eff} (\theta)$ can give the physical spectrum of the rational space-time crystal system, the topological classification of the rational space-time crystal system can be reduced to classifying $H_{eff} (\theta)$.  

      \subsection{Hermitian Case }
        For Hermitian rational space-time crystal system, the Hamiltonian $H$ in Eq.\eqref{8} is Hermitian. Thus, $\tilde{H}_{n} = \tilde{H}_{-n}^{\dagger}$. According to Eq.\eqref{21}, the effective Hamiltonian $H_{eff} (\theta)$ corresponding to $H$ is also Hermitian.
        \par

        There are $3$ types for Hermitian rational space-time crystal systems:
          \begin{Definition}
             Gapped --- The Hermitian effective Hamiltonian under PBC, $H_{eff}(\mathbf{k},\theta)$, is gapped for all $\theta \in [0,2\pi]$.
          \end{Definition}
          \begin{Definition}
            Half-gapped ---  The Hermitian effective Hamiltonian under PBC, $H_{eff}(\mathbf{k},\theta)$, is gapless for some specific values of $\theta$.
          \end{Definition}
          \begin{Definition}
            Gapless ---  The Hermitian effective Hamiltonian under PBC, $H_{eff}(\mathbf{k},\theta)$, is gapless for all $\theta \in [0,2\pi]$.
          \end{Definition}
         Here, we discuss the topological classification of the gapped systems. Since we have obtained the representation of internal symmetries on the effective Hamiltonian, a Hermitian $H_{eff}(\mathbf{k},\theta)$ can be classified by tenfold AZ class~\cite{10}. Since $\theta$ is not a momentum-like variable, we treat $H_{eff}(\mathbf{k},\theta)$ as a $d$-dimensional Hamiltonian if $\mathbf{k}$ is a $d$-dimensional momentum. Thus, the topological invariant about $H_{eff}(\mathbf{k},\theta)$, $\mathbf{inv}(\theta)$, given by the traditional tenfold classification~\cite{10}, contains $\theta$. For gapped system, the gap of $H_{eff}(\mathbf{k},\theta)$ will not be closed when $\theta$ varies from $0$ to $2\pi$. This means that if we treat $H_{eff}(\mathbf{k},\theta)$ as a $d$-dimensional Hamiltonian with a parameter $\theta$, $H_{eff}(\mathbf{k},\theta)$ experience no topological transition when $\theta$ changes from $0$ to $2\pi$. Hence, $\mathbf{inv}(\theta)$ is well defined for all $\theta \in [0,2\pi]$, and for arbitrary $\theta_1, \theta_2 \in [0,2\pi]$, $\mathbf{inv}(\theta_1) = \mathbf{inv}(\theta_2)$. Thus, $\mathbf{inv}(\theta)$ is a good topological invariant for gapped rational Hermitian space-time crystal system.

      \subsection{Non-Hermitian case}
         For non-Hermitian rational space-time crystal system, the effective Hamiltonian is no longer Hermitian. For a given $\theta_0 \in [0,2\pi]$, according to Ref.\cite{8}, there are $3$ types of the complex energy gap, which are line gap, point gap and gapless. Hence, there are $3$ types for non-Hermitian rational space-time systems:
         
           \begin{Definition}
             Gap-preserving --- The non-Hermitian effective Hamiltonian, under PBC $H_{eff}(\mathbf{k},\theta)$, has line gap for all $\theta \in [0,2\pi]$, or $H_{eff}(\mathbf{k},\theta)$ has point gap for all $\theta \in [0,2\pi]$.
           \end{Definition}
           \begin{Definition}
             Gap-breaking --- For non-Hermitian effective Hamiltonian under PBC, $H_{eff}(\mathbf{k},\theta)$, there exist $\theta_1,\theta_2 \in [0,2\pi]$ and $\theta_1 \not= \theta_2$, such that the gap type of $H_{eff}(\mathbf{k},\theta_1)$ and $H_{eff}(\mathbf{k},\theta_2)$ are different.  
           \end{Definition}
           \begin{Definition}
              Gapless --- The non-Hermitian effective Hamiltonian under PBC, $H_{eff}(\mathbf{k},\theta)$, is gapless for all $\theta \in [0,2\pi]$.
           \end{Definition}
         In this subsection, we discuss the topological classification of gap-preserving systems. Similar to the Hermitian case, the non-Hermitian rational space-time crystal system can be classified by 38-fold classification of the effective Hamiltonian $H_{eff}(\mathbf{k},\theta)$ and the type of energy gap according to Ref.\cite{8}. Thus, for $H_{eff}(\mathbf{k},\theta)$ with $d$-dimensional $\mathbf{k}$, the topological invariant of $H_{eff}(\mathbf{k},\theta)$, $\mathbf{inv}(\theta)$, is the topological invariant given by Ref.\cite{8} for $d$-dimensional non-Hermitian system, which contains a parameter $\theta$. For gap-preserving system, the energy gap is always preserved when $\theta$ changes form $0$ to $2\pi$. This means, if we treat $H_{eff}(\mathbf{k},\theta)$ as $d$-dimensional Hamiltonian with a parameter $\theta$, this Hamiltonian will not experience a topological transition when $\theta$ varies from $0$ to $2\pi$. Thus, $\mathbf{inv}(\theta)$ is well defined for all $\theta \in [0,2\pi]$ for gap-preserving non-Hermitian rational space-time crystal system, and the value of $\mathbf{inv}(\theta)$ is independent of $\theta$. We choose $\mathbf{inv}(\theta)$ as the topological invariant of gap-preserving non-Hermitian rational space-time crystal system.

  \section{Examples}
      \subsection{Hermitian case}
            Consider a $(1+1)$-dimensional space-time crystal system with two orbitals. The lattice potential has the form
              \begin{equation}
                V(r,t) = V_0 (r) + V_1 (r,t),
                \label{DD1}
              \end{equation}
            with
              \begin{equation}
                V_0 (r) = V_0 (r+a), \qquad V_1 (r,t) \propto cos(\frac{\pi}{a} r - \frac{2 \pi}{T} t),
                \label{DD2}
              \end{equation}
            where $a$ is the lattice constant. Thus, this system has space-time translation symmetry,
              \begin{equation}
                V (r,t) = V (r+a,t+ \frac{T}{2}), \quad V (r,t) = V (r+2a,t+T).
                \label{DD3}
              \end{equation}
            Apparently, this system is a rational space-time crystal system. We choose 
              \begin{equation}
                \mathbf{a}_0 = (0,-T), \qquad \mathbf{a}_1 = (a, \frac{T}{2})
                \label{DD4}
              \end{equation}
            as basis of discrete space-time translation symmetry. The reciprocal vectors of them are~\footnotemark[2]
              \begin{equation}
                 \mathbf{b}_0 = (\frac{\pi}{a},\frac{2\pi}{T}), \qquad  \mathbf{b}_1 = (\frac{2\pi}{a},0).
              \end{equation}
            Now, we take $a=1$ and $T=\frac{2\pi}{5}$. Then, we have $\delta k = \pi$ and $\Omega = 5$ for this system. The tight-binding Hamiltonian of this system is given by Eqs.\eqref{6} and \eqref{7}. According to Ref.\cite{2}, for such a lattice potential given by Eqs.\eqref{DD1} and \eqref{DD2}, only $H_{-1}$, $H_0$ and $H_1$ are nonzero. Hence, we assume that,
              \begin{equation}
                H_0 = \sum_n t_{0}^{(0)} c_{n,A}^{\dagger} c_{n,B} + t_{1}^{(0)} c_{n,B}^{\dagger} c_{n+1,A} + H.c.    
                \label{D1}
              \end{equation}
              \begin{equation}
                H_1 = \sum_n t_{1}^{(1)} e^{i n \pi} c_{n,A}^{\dagger} c_{n,B},
                \label{D2}
              \end{equation}
            and
              \begin{equation}
                H_{-1} = H_1^{\dagger},
                \label{D3}
              \end{equation}
            where $A$ and $B$ are indexes of orbitals and $n \in [1,L]$ is the index of site and $L$ is the length of the system. In matrix form,
              \begin{equation}
                H_0 = 
                   \begin{pmatrix}
                      0 & t_{0}^{(0)} &  &  &   &   \\
                      t_{0}^{(0)} & 0 & t_{1}^{(0)}  &  &   &   \\
                      & t_{1}^{(0)} & 0 & t_{0}^{(0)} & &   \\
                      & & t_{0}^{(0)} & 0 & t_{1}^{(0)} &   \\
                      & & & t_{1}^{(0)} & 0 & \ddots    \\
                      & & & & \ddots & \ddots 
                   \end{pmatrix}_{2L \times 2L},
                   \label{D4}
              \end{equation}
           and
              \begin{equation}
                 H_1 =  
                   \begin{pmatrix}
                      0 & -t_{1}^{(1)} & & &  &   \\
                      & 0 & 0 & &  &  & \\
                      & & 0 & t_{1}^{(1)} &  &   \\
                      & & & 0 & 0 &    \\
                      & & & & \ddots & \ddots  \\
                      & & & & & \ddots
                   \end{pmatrix}_{2L \times 2L} .
                   \label{D5}
              \end{equation}
            Thus, According to Eq.(8) and Eq.(21), 
              \begin{equation}
                H_{eff} (\theta) =
                   \begin{pmatrix}
                      \mathbf{h}_0  &  \mathbf{h}_1  & & &  \\
                      \mathbf{h}_1^{\dagger} & \mathbf{h}_0 & \mathbf{h}_1 & & \\
                      & \mathbf{h}_1^{\dagger} & \mathbf{h}_0 & \mathbf{h}_1 &  \\
                      & & \ddots & \ddots & \ddots  \\
                      & & & \ddots & \ddots
                   \end{pmatrix}_{2L \times 2L}   ,
                   \label{D6}
              \end{equation}
            where
              \begin{multline}
                \mathbf{h}_0 = \\
                  \begin{pmatrix}
                   0 & t_{0}^{(0)} - t_{1}^{(1)} e^{i \theta} & 0 & 0  \\
                   t_{0}^{(0)} - t_{1}^{(1)} e^{-i \theta} & 0 &  t_{1}^{(0)} & 0  \\
                   0 & t_{1}^{(0)} & 0 & t_{0}^{(0)} + t_{1}^{(1)} e^{i \theta} \\
                   0 & 0 & t_{0}^{(0)} + t_{1}^{(1)} e^{-i \theta} & 0 
                  \end{pmatrix} ,
                  \label{D7}
              \end{multline}
             and 
               \begin{equation}
                  \mathbf{h}_1 = 
                     \begin{pmatrix}
                        0 & 0 & 0 & 0 & \\
                        0 & 0 & 0 & 0  &\\
                        0 & 0 & 0 & 0  &\\
                        t_{1}^{(0)} & 0 & 0 & 0 & 
                     \end{pmatrix} .
                     \label{D8}
               \end{equation}
            Under PBC,
              \begin{multline}
                H_{eff}(k,\theta) = \mathbf{h}_0 + \mathbf{h}_1 e^{-i k} + \mathbf{h}_1^{\dagger} e^{i k}  =  \\
                  \begin{pmatrix}
                   0 & t_{0}^{(0)} - t_{1}^{(1)} e^{i \theta} & 0 & t_{1}^{(0)} e^{i k}  \\
                   t_{0}^{(0)} - t_{1}^{(1)} e^{-i \theta} & 0 &  t_{1}^{(0)} & 0 \\
                   0 & t_{1}^{(0)} & 0 & t_{0}^{(0)} + t_{1}^{(1)} e^{i \theta}  \\
                   t_{1}^{(0)} e^{-i k} & 0 & t_{0}^{(0)} + t_{1}^{(1)} e^{-i \theta} & 0 
                  \end{pmatrix}.
                  \label{D9}
              \end{multline}
            This system has chiral symmetry,
              \begin{equation}
                \mathcal{S} H_{eff}(\mathbf{k},\theta)  \mathcal{S}^{-1} = -H_{eff}(\mathbf{k},\theta),
                \label{D10}
              \end{equation}
            where $ \mathcal{S} = \mathbf{I}_{2 \times 2} \otimes \sigma_z$, $ \mathbf{I}_{2 \times 2}$ is the $2$-dimensional identity matrix and $\sigma_z$ is the Pauli matrix. Thus this system is $\mathbf{AIII}$ class.
            \par
 
\begin{figure}
                \centering
                 \subfigure[]{\includegraphics[scale=0.31]{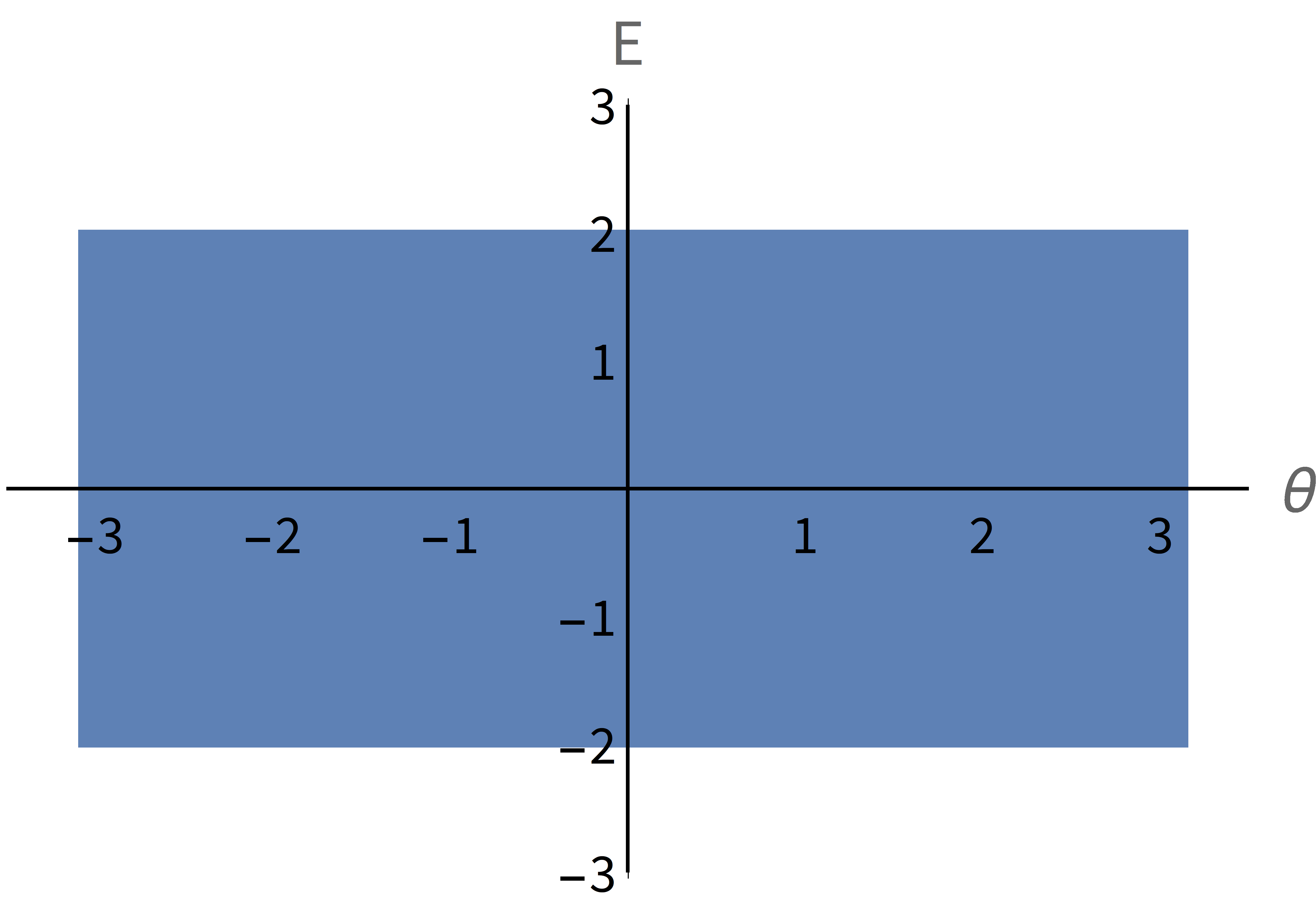} \label{fig1a}}
                 \quad
                 \subfigure[]{\includegraphics[scale=0.31]{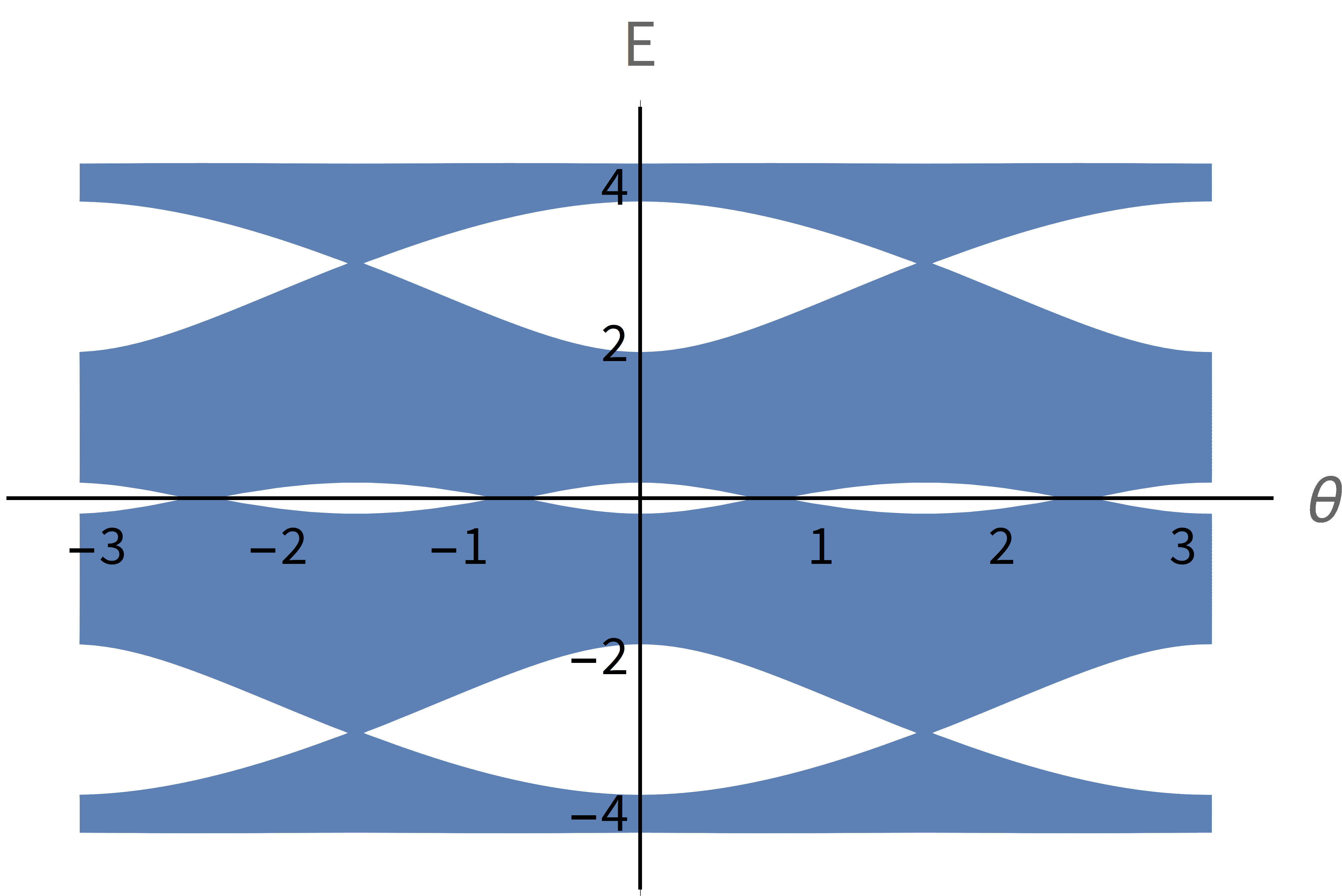} \label{fig1b}}
                \caption{(a).The blue area is the spectrum of the effective Hamiltonian $H_{eff}(k,\theta)$ given by Eq.\eqref{D9} with $t_{0}^{(0)} = 0$, $t_{1}^{(0)} = 1$ and $t_{1}^{(1)}=1$ under PBC. (b).The blue area is the spectrum of the effective Hamiltonian $H_{eff}(k,\theta)$ given by Eq.\eqref{D9} with $t_{0}^{(0)} = 2$, $t_{1}^{(0)} = 2$ and $t_{1}^{(1)}=1$ under PBC.  }
                \label{fig1}
              \end{figure}

            For the case $t_{0}^{(0)} = 0$, $t_{1}^{(0)} = 1$ and $t_{1}^{(1)}=1$, the system is gapless~(Fig.\ref{fig1a}), and for the case $t_{0}^{(0)} = 2$, $t_{1}^{(0)} = 2$ and $t_{1}^{(1)}=1$, the system is half-gapped~(Fig.\ref{fig1b}).
            \par
 
\begin{figure}
                \centering
                \subfigure[]{\includegraphics[scale=0.31]{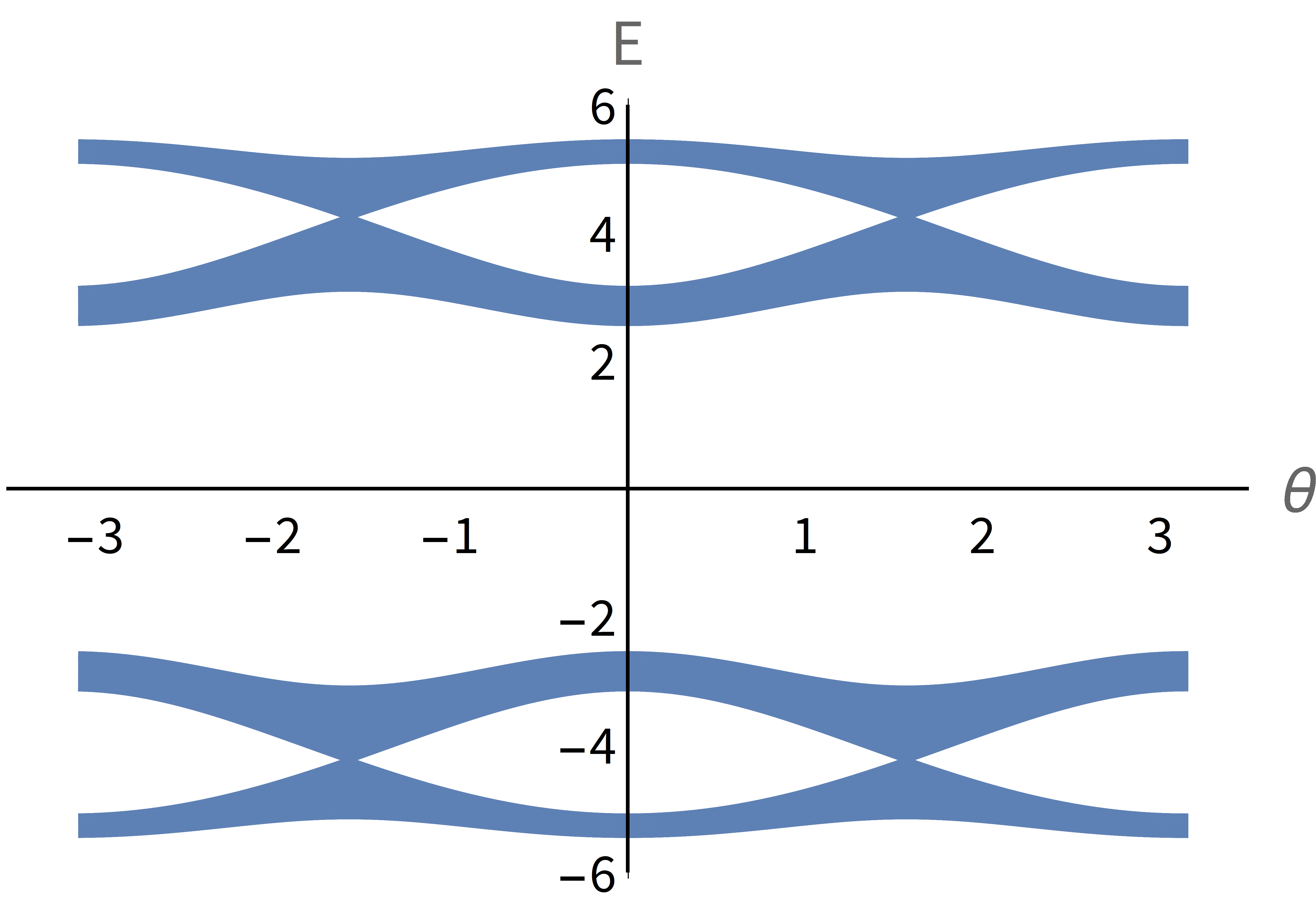} \label{fig2a}}
                \quad
                \subfigure[]{\includegraphics[scale=0.31]{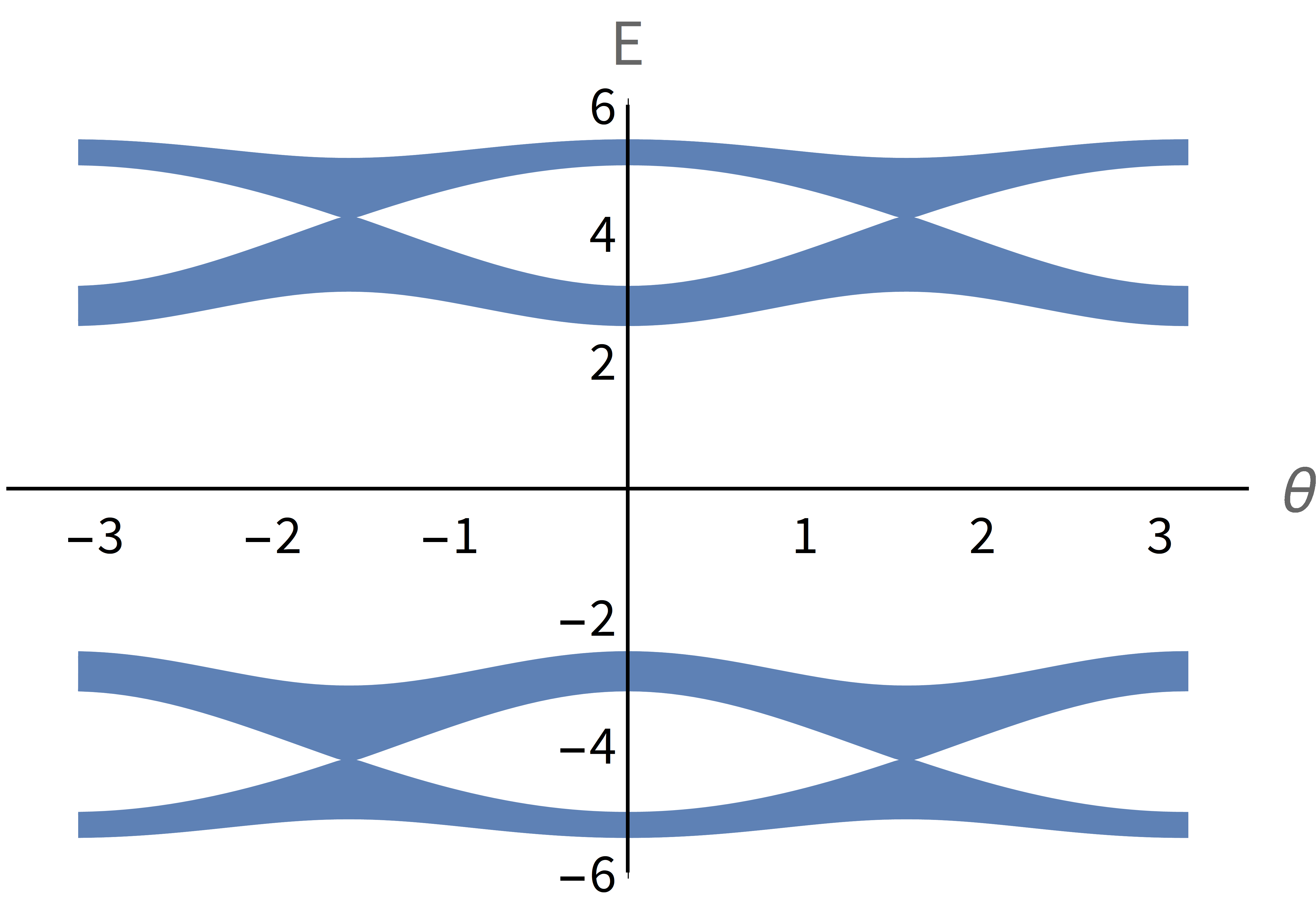} \label{fig2b}}
                \quad
                \subfigure[]{\includegraphics[scale=0.5]{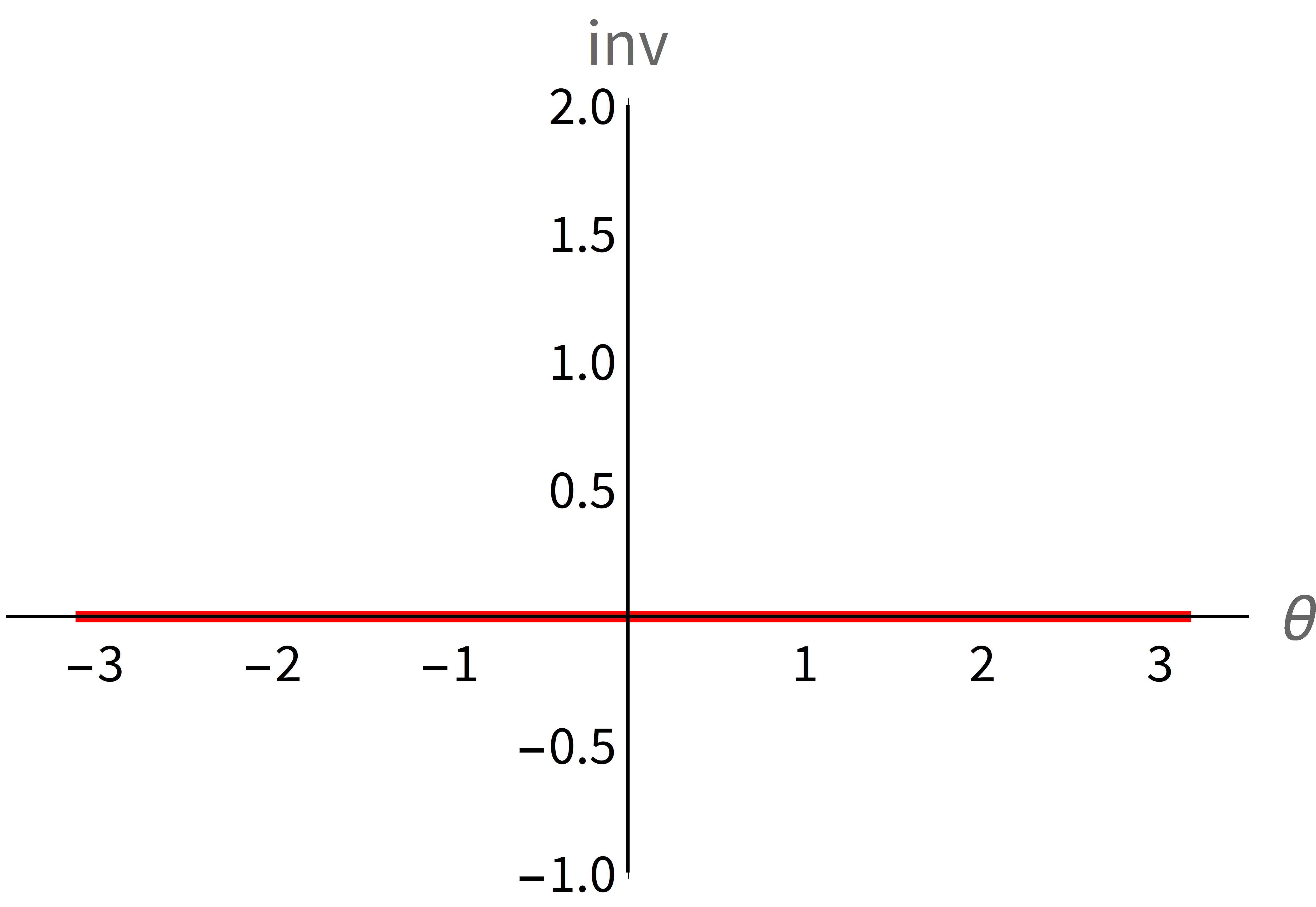} \label{fig2c}}
                \caption{For the case $t_{0}^{(0)} = 4$, $t_{1}^{(0)} = 1$ and $t_{1}^{(1)}=1$, (a).The blue area is the spectrum of the effective Hamiltonian $H_{eff}(k,\theta)$ given by Eq.\eqref{D9} under PBC.  (b).The blue area is the spectrum of he effective Hamiltonian $H_{eff}(\theta)$ given by Eq.\eqref{D6} under OBC with $L=50$.  (c).The red line is the winding number $\mathbf{inv}(\theta)$ of this this system.}
                \label{fig2}
              \end{figure}

\begin{figure}
                \centering
                \subfigure[]{\includegraphics[scale=0.31]{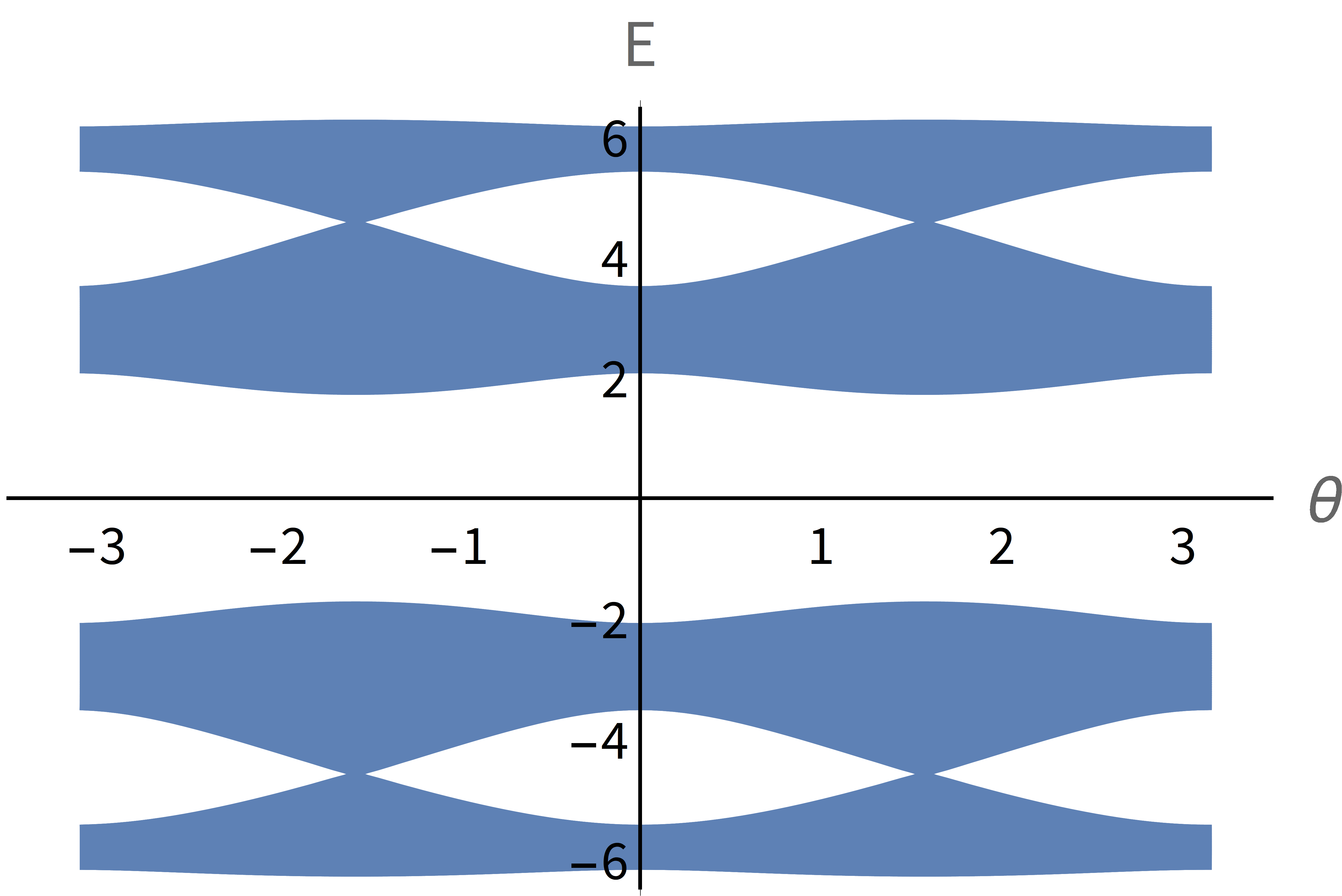} \label{fig3a}}
                \quad
                \subfigure[]{\includegraphics[scale=0.31]{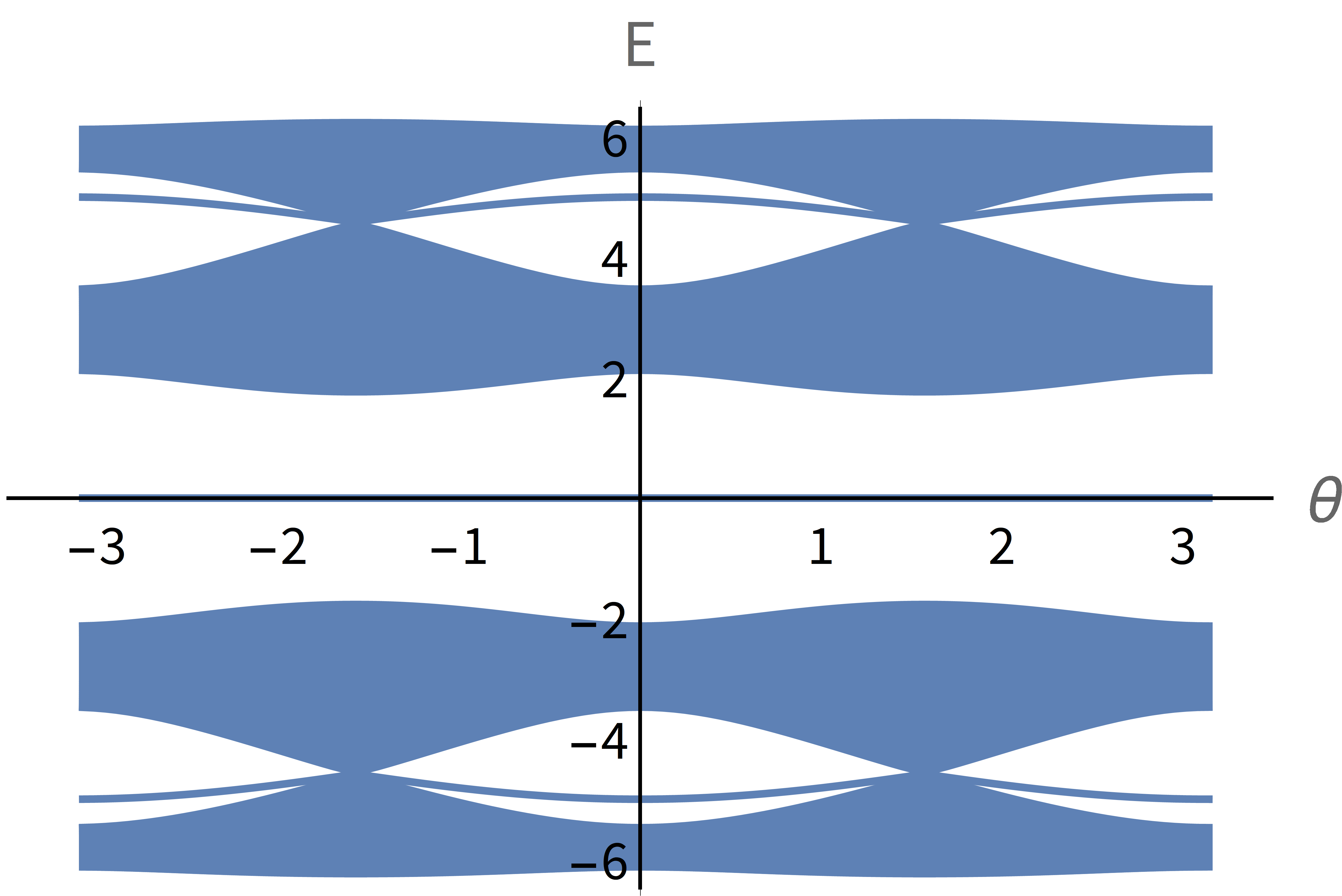} \label{fig3b}}
                \quad
                \subfigure[]{\includegraphics[scale=0.5]{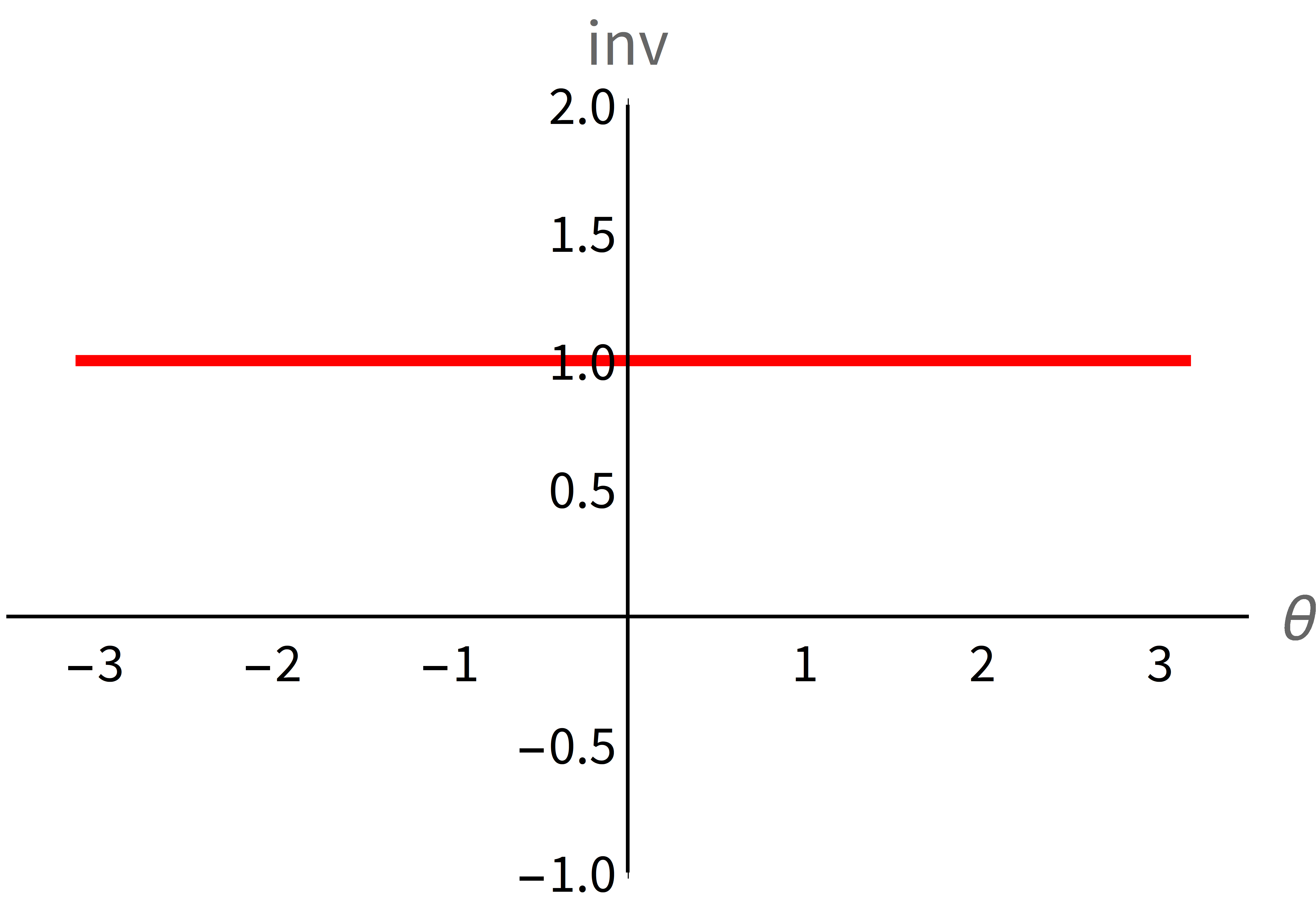} \label{fig3c}}
                \caption{For the case $t_{0}^{(0)} = 2$, $t_{1}^{(0)} = 4$ and $t_{1}^{(1)}=1$, (a).The blue area is the spectrum of the effective Hamiltonian $H_{eff}(k,\theta)$ given by Eq.\eqref{D9} under PBC.  (b).The blue area is the spectrum of he effective Hamiltonian $H_{eff}(\theta)$ given by Eq.\eqref{D6} under OBC with $L=50$.  (c).The red line is the winding number $\mathbf{inv}(\theta)$ of this this system.}
                \label{fig3}
              \end{figure}

            If $H_{eff}(k,\theta)$ give by Eq.\eqref{D9} is gapped, the topological invariant of system belonged to $\mathbf{AIII}$ class is winding number,
              \begin{equation}
                \mathbf{inv}(\theta) = \frac{i}{2(2\pi)} \int_{-\pi}^{\pi}  Tr[ \mathcal{S} H_{eff}^{-1}(k,\theta) \frac{d}{dk} H_{eff}(k,\theta) ] \mathrm{d}k.
                \label{D11}
              \end{equation}
            If we take $t_{0}^{(0)} = 4$, $t_{1}^{(0)} = 1$ and $t_{1}^{(1)}=1$, the system is gapped and topological trivial. In this case, $\mathbf{inv}(\theta)=0$ for all $\theta$~(Fig.\ref{fig2}). For the case $t_{0}^{(0)} = 2$, $t_{1}^{(0)} = 4$ and $t_{1}^{(1)}=1$, this system is gapped and topological non-trivial and $\mathbf{inv}(\theta)=1$ for all $\theta$~(Fig.\ref{fig3}). For topological non-trivial system, there exist topological state in the spectrum under OBC~(Fig.\ref{fig3b}). For topological trivial system, there exist no topological state in the OBC spectrum~(Fig.\ref{fig2b}). This is consistent with the traditional Hermitian bulk-boundary correspondence.

      \subsection{non-Hermitian case }
            We consider a $(1+1)$-dimensional non-Hermitian space-time crystal system with single orbital. For simplicity, we still assume the the lattice potential has the form given by Eqs.\eqref{DD1} and \eqref{DD2} with $a=1$ and $T=\frac{2\pi}{5}$. Thus, $\delta k = \pi$ and $\Omega = 5$ for this system, and only $H_{-1}$, $H_0$ and $H_1$ are nonzero for the tight-binding Hamiltonian given by Eq.\eqref{6}. Hence, we assume that,
              \begin{gather}
                 H_0 = \sum_n t_{1}^{(0)} c_{n+1}^{\dagger} c_{n} + t_{-1}^{(0)} c_{n-1}^{\dagger} c_{n}   \label{D12} \\
                 H_1 = \sum_n \gamma e^{i(n+1)\pi} c_{n+1}^{\dagger} c_{n} + \gamma e^{i(n-1)\pi} c_{n-1}^{\dagger} c_{n}  \label{D13}    \\
                 H_{-1} = \sum_n \gamma e^{i(n+1)\pi} c_{n+1}^{\dagger} c_{n} + \gamma e^{i(n-1)\pi} c_{n-1}^{\dagger} c_{n} , \label{D14}
              \end{gather}
            where $n \in [1,L]$ is the index of site and $L$ is the length of the system. Under PBC, according to Eq.(8) and Appendix A,
               \begin{gather}
                 \tilde{H}_0 = 
                   \begin{pmatrix}
                     0 & t_{-1}^{(0)} + t_{1}^{(0)} e^{i k}  \\
                     t_{1}^{(0)} + t_{-1}^{(0)} e^{-i k}  & 0
                   \end{pmatrix}    \label{D15}
                   \\
                  \tilde{H}_1 =
                    \begin{pmatrix}
                      0 & -\gamma - \gamma e^{i k}  \\
                      \gamma + \gamma e^{-i k}  & 0
                    \end{pmatrix}  \label{D16}
                    \\
                  \tilde{H}_{-1} =
                     \begin{pmatrix}
                      0 & -\gamma - \gamma e^{i k}  \\
                      \gamma + \gamma e^{-i k}  & 0
                     \end{pmatrix}.  \label{D17}
               \end{gather}
            Thus, the effective Hamiltonian under PBC is 
               \begin{equation}
                  \begin{split}
                   H_{eff}(k,\theta)  &= \tilde{H}_0 + \tilde{H}_{1} e^{i \theta} + \tilde{H}_{-1} e^{-i \theta}  \\
                   & = 
                    \begin{pmatrix}
                       0 & f_1 (k,\theta)   \\
                       f_2 (k,\theta) & 0
                    \end{pmatrix},
                  \end{split}
                  \label{D18}
               \end{equation}
            where
               \begin{gather}
                 f_1 (k,\theta) = (t_{-1}^{(0)} - 2\gamma cos(\theta)) + (t_{1}^{(0)} - 2\gamma cos(\theta)) e^{i k}    
                 \label{D19}
                 \\
                 f_2 (k,\theta) = (t_{1}^{(0)} + 2\gamma cos(\theta)) + (t_{-1}^{(0)} + 2\gamma cos(\theta)) e^{-i k} .
                 \label{D20}
               \end{gather}
            We find that
               \begin{equation}
                  \Gamma H_{eff}^{\dagger}(k,\theta) \Gamma^{-1} = - H_{eff}(k,\theta),
                  \label{D21}
               \end{equation}
            with $\Gamma = \sigma_z K P$, where $\sigma_z$ is the Pauli matrix, $K$ is the complete conjugation operator and $P$ is the operator acting on creation and annihilation operators,
               \begin{equation}
                 P c_{k}^{\dagger} P^{-1} = c_{k} , \qquad
                 P c_{k} P^{-1} = c_{k}^{\dagger}.
                 \label{D22}
               \end{equation} 
            Thus, this system belongs to $\mathbf{A}$ class with sublattice symmetry. For $ H_{eff}(k,\theta)$ given in Eq.\eqref{D18}, the eigenenergy is
               \begin{multline}
                  E^{2} (k,\theta) = - Det(H_{eff}(k,\theta)) = f_1 (k,\theta) f_2 (k,\theta) \\
                  = e^{-i k} [ (e^{i k} t_{1}^{(0)} + t_{-1}^{(0)} )^2 - 4 (1+ e^{i k})^2 \gamma^2 cos^2 (\theta)     ].
                  \label{D23}
               \end{multline}

               \begin{figure}
                 \centering
                 \includegraphics[scale=0.5]{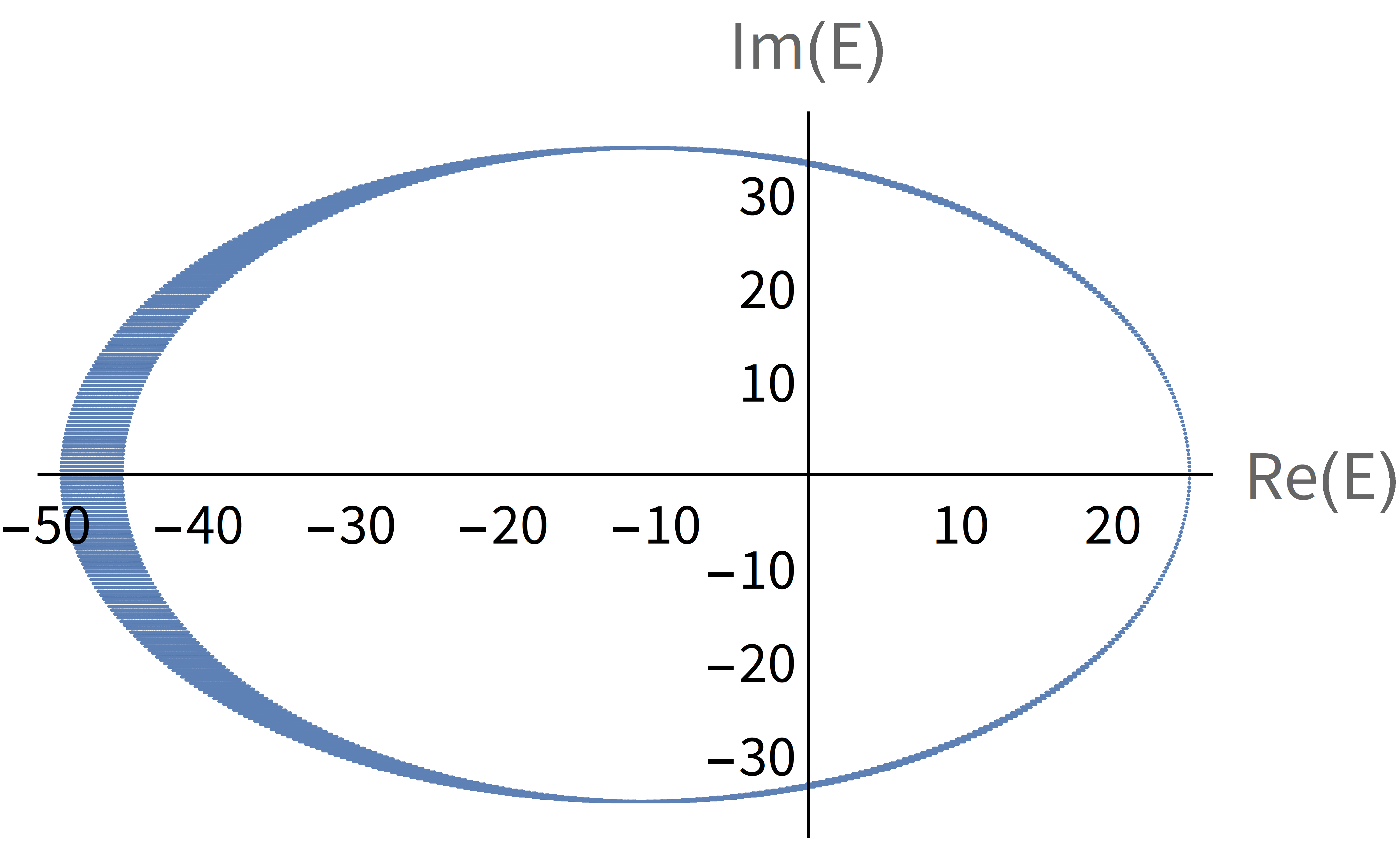} 
                 \caption{The blue area is the values of $E^{2} (k,\theta)$ with $t_{1}^{(0)} =1$, $t_{-1}^{(0)} =6$ and $\gamma = \frac{1}{2}$ for $k,\theta \in [0.2\pi]$. }
                 \label{fig4}
               \end{figure}
               \par
             
            For the case $t_{1}^{(0)} =1$, $t_{-1}^{(0)} =6$ and $\gamma = \frac{1}{2}$, $E^{2} (k,\theta) = - Det(H_{eff}(k,\theta)) \not= 0 $ for all $k,\theta \in [0,2\pi]$~(Fig.\ref{fig4}). According to the definition $1$ in Ref.\cite{8}, $H_{eff}(k,\theta)$ has point gap for all $\theta \in [0,2\pi]$. Thus, this system is gap-preserving if $t_{1}^{(0)} =1$, $t_{-1}^{(0)} =6$ and $\gamma = \frac{1}{2}$. In this case, this system has two topological invariants~\cite{8},
               \begin{gather}
                 \mathbf{inv}_1 (\theta) = \frac{-i}{2\pi} \int_{-\pi}^{\pi} f_{1}^{-1} (k,\theta) \frac{d}{dk} f_{1} (k,\theta) \mathrm{d}k
                 \label{D24}   \\
                 \mathbf{inv}_2 (\theta) = \frac{-i}{2\pi} \int_{-\pi}^{\pi} f_{2}^{-1} (k,\theta) \frac{d}{dk} f_{2} (k,\theta) \mathrm{d}k
                 \label{D25}.
               \end{gather} 
            \par

            If $t_{1}^{(0)} = t_{-1}^{(0)} $, $E^{2} (\pi,\theta) = 0$ for arbitrary $\theta \in [0,2\pi]$. Thus, this system is gapless in this case.
            \par
 
            If we take $t_{1}^{(0)} =-1$, $t_{-1}^{(0)} =3$ and $\gamma = \frac{1}{2}$,
              \begin{equation}
                E^{2} (k,\theta) = e^{-i k} [ (-e^{i k}  + 3 )^2 -  (1+ e^{i k})^2  cos^2 (\theta)     ].
                \label{D26}
              \end{equation}
            If $\theta = \frac{\pi}{2}$, $E^{2} (k,\frac{\pi}{2}) = e^{-i k} (-e^{i k}  + 3 )^2 \not= 0 $ for all $k \in [-\pi,\pi]$. If $\theta = 0$, $E^{2} (k,0) = e^{-i k} [ (-e^{i k}  + 3 )^2 -  (1+ e^{i k})^2]$ and $E^{2} (0,0) = 0$. Hence, $H_{eff}(k,\frac{\pi}{2})$ has point gap and $H_{eff}(k,0)$ is gapless. This means this system is gap-breaking for the case $t_{1}^{(0)} =-1$, $t_{-1}^{(0)} =3$ and $\gamma = \frac{1}{2}$.

      \section{Conclusion and Discussion}
        In this work, we show that space-time crystal system can be divided into two classes, rational space-time crystal system and irrational space-time crystal system. A rational space-time crystal system is equal to a traditional floquet system. Then we find a way to solve the floquet equation analytically. Hence, we can obtain the quasi-spectrum and the effective Hamiltonian of the system analytically, and we point out that the physical spectrum of the system is just the $0$th sector of the quasi-spectrum. Then, we discuss the properties of the effective Hamiltonian under internal symmetries and give the types of the Hermitian and non-Hermitian effective Hamiltonian according to the energy gap. There are three types of Hermitian effective Hamiltonian, gapped, half-gapped and gapless. And there are also three types of non-Hermitian effective Hamiltonian, gap-preserving, gap-breaking and gapless. For gapped Hermitian rational space-time crystal systems, the topological classification of them can be reduced to the Hermitian tenfold AZ classes, and for gap-preserving non-Hermitian rational space-time crystal system under PBC, the topological classification of them can be reduced to the non-Hermitian $38$-fold classes with three types of the complex energy gap. Our works give a new perspective of space-time crystal system and pave a way to research rational space-time crystal system systematically.
        \par

        In the future, the topological classification about the half-gapped and gapless Hermitian rational space-time crystal systems and the topological classification about the gap-breaking and gapless non-Hermitian rational space-time crystal systems will be considered for the complete topological classification of rational space-time crystal systems. Furthermore, the properties of irrational space-time crystal systems will also be discussed in our future work .

    \section{acknowledgements}
       The authors thank useful discussion with Yongxu Fu, Haoshu Li and Zhiwei Yin. This work was supported by
       NSFC Grant No. 11275180.

  \appendix

     \section{The Hamiltonian After Taking the New Unit Cell}
       Consider a site locating at $\mathbf{R}$. Under the basis $\{ \mathbf{s}_i \}$, we can decompose $\mathbf{R}$ as $\mathbf{R} = \sum_{i=1}^d r_i \mathbf{s}_i$ with $r_i \in \mathbb{Z}$. Thus, we can use a column vector to represent $\mathbf{R}$,
         \begin{equation}
            \mathbf{R} = (r_1, r_2, \cdots, r_d)^T.
            \label{A1}
         \end{equation}
        The integer $r_i$ can be decomposed to $r_i = l_i q_i + h_i$ with $l_i, h_i \in \mathbb{Z}$ and $0 \leq h_i < q_i$. Then, $\mathbf{R}$ can be rewritten as
         \begin{equation}
            \mathbf{R} = (l_1 q_1 + h_1, l_2 q_2 + h_2, \cdots, l_d q_d +h_d)^T.
            \label{A2}
         \end{equation}
        After taking $\{ q_i \mathbf{s}_i \}$ as the group of lattice vectors, there are $\prod_{i=1}^d q_i$ orbitals in a unit cell. In this case, we treat the site locating at $\mathbf{R}$ under original coordinate as an orbital locating at 
         \begin{equation}
            \tilde{\mathbf{R}} = (l_1, l_2, \cdots, l_d)^T
            \label{A3}
         \end{equation} 
        with orbital indexes $(h_1, h_2, \cdots, h_d)$. Then, we define
         \begin{equation}
            \mathbf{c}_{\tilde{\mathbf{R}}} = (c_{\tilde{\mathbf{R}}}^{(0,0,\cdots,0)}, c_{\tilde{\mathbf{R}}}^{(1,0,\cdots,0)}, \cdots, c_{\tilde{\mathbf{R}}}^{(q_1-1,q_2-1,\cdots,q_d-1)})^T
            \label{A4}
         \end{equation} 
        as the annihilation operator about the new unit cell locating at $\tilde{R}$ and $(0, 0, \cdots, 0), (1,0,\cdots,0), \cdots, (q_1-1,q_2-1,\cdots,q_d-1) $ are  orbital indexes. For
          \begin{align}
            \mathbf{R}_i = \tilde{\mathbf{R}}_i + \sum_{m=1}^d h_{i,m} \mathbf{s}_m    \label{A5}  
            \\
            \mathbf{R}_j = \tilde{\mathbf{R}}_j + \sum_{m=1}^d h_{j,m} \mathbf{s}_m  ,  \label{A6}
          \end{align}
        we define
          \begin{multline}
            (\mathbf{M}_{\tilde{\mathbf{R}}_i - \tilde{\mathbf{R}}_j}^{(n)})_{(h_{i,1}, h_{i,2}, \cdots, h_{i,d}), (h_{j,1}, h_{j,2}, \cdots, h_{j,d})}  \\
            = e^{i n  \mathbf{ \delta k} \cdot \mathbf{R}_i} t_{\mathbf{R}_i-\mathbf{R}_j}^{(n)}.
            \label{A7}
          \end{multline}
         For $\tilde{\mathbf{R}} = (l_1, l_2, \cdots, l_d)^T$, $\mathbf{\delta k} \cdot \tilde{\mathbf{R}} = 2 \pi (\sum_{i=1}^d p_i l_i)$. Thus, $e^{ i n \mathbf{\delta k} \cdot \tilde{\mathbf{R}}} = 1$, and 
           \begin{equation}
            e^{i n  \mathbf{ \delta k} \cdot \mathbf{R}_i} = e^{i n  \mathbf{ \delta k} \cdot (\sum_{m=1}^d h_{i,m} \mathbf{s}_m)}.
            \label{A8}
           \end{equation}
        This means the matrix $\mathbf{M}_{\tilde{\mathbf{R}}_i - \tilde{\mathbf{R}}_j}^{(n)}$ only depends on the value of $\tilde{\mathbf{R}}_i - \tilde{\mathbf{R}}_j$. Hence, the Hamiltonian in Eq.(6) and Eq.(7) can be written as the Hamiltonian in Eq.(8), which respect the spatial translation symmetry.

      \section{The Physical Spectrum of The System}
        For the effective Hamiltonian $H_{eff}(\theta)$, we use $| u_{\omega}^{\theta} \rangle$ to represent the eigenvector of $H_{eff}(\theta)$ with eigenvalue $\omega$. For a given $\theta$, $\{ | u_{\omega}^{\theta} \rangle \}$ are orthogonal and complete,
           \begin{equation}
             \langle u_{\omega^{\prime}}^{\theta} | u_{\omega}^{\theta} \rangle =\delta_{\omega^{\prime},\omega}  ,
             \qquad
             \sum_{\omega} | u_{\omega}^{\theta} \rangle \langle u_{\omega}^{\theta} | = \mathbf{1}.
             \label{B1}
           \end{equation}
         According to Eq.(10) and Eq.(12), the physical state corresponding to the quasistate $| u_{\omega}^{\theta} \rangle$ is
           \begin{eqnarray}
             | \psi_{\omega}^{\theta} (t) \rangle = \lim_{l \rightarrow \infty} e^{-i \omega t} \frac{1}{\sqrt{2l}} \sum_{m = -l}^{l} e^{- i m \Omega t} e^{- i m \theta} | u_{\omega}^{\theta} \rangle .
             \label{B2}
           \end{eqnarray}
         The expected value of energy about the state $| \psi_{\omega}^{\theta} (t) \rangle$ under long time average is 
           \begin{equation}
              \begin{split}
                \bar{E} & = \lim_{t_0 \rightarrow \infty} \frac{1}{2 t_0} \int_{-t_0}^{t_0} \langle \psi_{\omega}^{\theta} (t) | H(t) | \psi_{\omega}^{\theta} (t) \rangle \mathrm{d}t \\
                       & =  \lim_{t_0 \rightarrow \infty} \frac{1}{2 t_0} \int_{-t_0}^{t_0} \langle \psi_{\omega}^{\theta} (t) | i \partial_t | \psi_{\omega}^{\theta} (t) \rangle \mathrm{d}t
              \end{split}
              \label{B3}
           \end{equation}
         According to Eq.\eqref{B2},
           \begin{multline}
              \langle \psi_{\omega}^{\theta} (t) | i \partial_t | \psi_{\omega}^{\theta} (t) \rangle = \lim_{l \rightarrow \infty} \frac{1}{2l} \sum_{m = -l}^{l} \sum_{n = -l}^{l} (\omega + m \Omega)  \\
               \times e^{- i (m-n) \Omega t} e^{- i (m-n)\theta} \langle u_{\omega}^{\theta} | u_{\omega}^{\theta} \rangle .
               \label{B4}
           \end{multline}
         Since
            \begin{gather}
               \langle u_{\omega}^{\theta} | u_{\omega}^{\theta} \rangle = 1 
                 \label{B5}  \\
               \lim_{t_0 \rightarrow \infty} \frac{1}{2 t_0} \int_{-t_0}^{t_0} e^{- i (m-n) \Omega t} {d}t = \delta_{m,n} ,
               \label{B6}
            \end{gather}
         Eq.\eqref{B3} can be reduced as 
            \begin{equation}
               \begin{split}
                  \bar{E} & = \lim_{l \rightarrow \infty} \frac{1}{2l} \sum_{m = -l}^{l} \sum_{n = -l}^{l} (\omega + m \Omega)  \delta_{m,n} e^{- i (m-n)\theta} \\
                          & = \omega   .
               \end{split}
               \label{B7}
            \end{equation}
         This means the expected value of energy about the state $| \psi_{\omega}^{\theta} (t) \rangle$, which is the physical state corresponding to the quasitate $| u_{\omega}^{\theta} \rangle$ with quasienergy $\omega$ in $0$th sector of the quasi-spectrum, under long time average is still $\omega$. Hence, the $0$th sector of the quasi-spectrum gives the physical spectrum of the system.
         \par

         If $H_{eff} (\theta)$ is non-Hermitian, the eigenvectors of $H_{eff} (\theta)$ are biorthogonal and complete~\cite{6,7},
            \begin{equation}
               \langle u_{\omega^{\prime},L}^{\theta} | u_{\omega,R}^{\theta} \rangle =\delta_{\omega^{\prime},\omega},  
               \qquad
               \sum_{\omega} | u_{\omega,R}^{\theta} \rangle \langle u_{\omega,L}^{\theta} | = \mathbf{1} ,
               \label{B8}
            \end{equation} 
         where $R$~($L$) means this state is a right~(left) eigenstate of $H_{eff} (\theta)$. In this case,
            \begin{gather}
               | \psi_{\omega,R}^{\theta} (t) \rangle = \lim_{l \rightarrow \infty} e^{-i \omega t} \frac{1}{\sqrt{2l}} \sum_{m = -l}^{l} e^{- i m \Omega t} e^{- i m \theta} | u_{\omega,R}^{\theta} \rangle
               \label{B9}  \\
               | \psi_{\omega,L}^{\theta} (t) \rangle = \lim_{l \rightarrow \infty} e^{-i \omega t} \frac{1}{\sqrt{2l}} \sum_{m = -l}^{l} e^{- i m \Omega t} e^{- i m \theta} | u_{\omega,L}^{\theta} \rangle
               \label{B10},
            \end{gather}
         and
            \begin{equation}
               \begin{split}
                  \bar{E} & = \lim_{t_0 \rightarrow \infty} \frac{1}{2 t_0} \int_{-t_0}^{t_0} \langle \psi_{\omega,L}^{\theta} (t) | H(t) | \psi_{\omega,R}^{\theta} (t) \rangle \mathrm{d}t \\
                       & =  \lim_{t_0 \rightarrow \infty} \frac{1}{2 t_0} \int_{-t_0}^{t_0} \langle \psi_{\omega,L}^{\theta} (t) | i \partial_t | \psi_{\omega,R}^{\theta} (t) \rangle \mathrm{d}t \\
                      & = \omega .
               \end{split}
               \label{B11}
            \end{equation}
         Hence, for non-Hermitian rational space-time crystal system, the $0$th sector of the quasi-spectrum still gives the physical spectrum.

      \section{The Representation of Internal Symmetry on the Effective Hamiltonian}
        \subsection{For Hermitian case}
          Under PBC, the Hamiltonian in Eq.(8) is written as 
           \begin{equation}
             \begin{split}
               H & = \sum_{\mathbf{k}} \sum_{n=-\infty}^{\infty} e^{-i n \Omega t}      \tilde{H}_n (\mathbf{k})  \\
                 & = \sum_{\mathbf{k}} \sum_{n=-\infty}^{\infty} e^{-i n \Omega t} \sum_{ \tilde{\mathbf{R}}} e^{-i \mathbf{k} \cdot \tilde{\mathbf{R}}} \mathbf{c}_{\mathbf{k}}^{\dagger} \mathbf{M}_{\tilde{\mathbf{R}}}^{(n)} \mathbf{c}_{\mathbf{k}}.
                 \label{C1}
             \end{split}
           \end{equation} 
          Under time-reversal transformation, operator $\mathbf{c}_{\mathbf{k}}$ and unit imaginary number $i$ are transformed as
           \begin{equation}
            \mathcal{T} \mathbf{c}_{\mathbf{k}} \mathcal{T}^{-1} = U_{\mathcal{T}} \mathbf{c}_{-\mathbf{k}}  ,
            \qquad
            \mathcal{T} i \mathcal{T}^{-1} = -i  ,
            \label{C2}
           \end{equation}
          where $U_{\mathcal{T}}$ is a unitary matrix. Thus, under TRS, the Hamiltonian $H$ is transformed as 
           \begin{equation}
             \begin{split}
               H_{\mathcal{T}} & = \mathcal{T} H \mathcal{T}^{-1} = \sum_{\mathbf{k}} \sum_{n=-\infty}^{\infty} e^{i n \Omega t} \mathcal{T} \tilde{H}_n (\mathbf{k}) \mathcal{T}^{-1}   \\
                  & = \sum_{\mathbf{k}} \sum_{n=-\infty}^{\infty} e^{i n \Omega t} \sum_{ \tilde{\mathbf{R}}} e^{i \mathbf{k} \cdot \tilde{\mathbf{R}}} \mathbf{c}_{-\mathbf{k}}^{\dagger} U_{\mathcal{T}}^{\dagger} \mathbf{M}_{\tilde{\mathbf{R}}}^{(n)*} U_{\mathcal{T}} \mathbf{c}_{-\mathbf{k}}        \\
                  & = \sum_{\mathbf{k}} \sum_{n=-\infty}^{\infty} e^{- i n \Omega t} \tilde{H}_{n,\mathcal{T}} (-\mathbf{k}),
             \end{split}
             \label{C3}
           \end{equation}
          where 
            \begin{equation}
               \tilde{H}_{n,\mathcal{T}} (-\mathbf{k}) = \sum_{ \tilde{\mathbf{R}}} e^{i \mathbf{k} \cdot \tilde{\mathbf{R}}} \mathbf{c}_{-\mathbf{k}}^{\dagger} U_{\mathcal{T}}^{\dagger} \mathbf{M}_{\tilde{\mathbf{R}}}^{(-n)*} U_{\mathcal{T}} \mathbf{c}_{-\mathbf{k}},
               \label{C4}
            \end{equation}
          and $*$ means complex conjugation.  Thus, the effective Hamiltonian corresponding to $H_{\mathcal{T}}$ is 
            \begin{equation}
               \mathcal{T} H_{eff} (\mathbf{k},\theta) \mathcal{T}^{-1} = H_{eff,\mathcal{T}} (\mathbf{k},\theta) = \sum_{n=-\infty}^{\infty} e^{- i n \theta}  \tilde{H}_{-n,\mathcal{T}} (\mathbf{k}).
               \label{C5}
            \end{equation}
          If the system has TRS, $H = H_{\mathcal{T}}$. Hence, for such a system, $\tilde{H}_{n,\mathcal{T}} (- \mathbf{k}) = \tilde{H}_{n} (\mathbf{k})$. Thus,
            \begin{equation}
               H_{eff,\mathcal{T}} (\mathbf{k},\theta) = \sum_{n=-\infty}^{\infty} e^{- i n \theta}  \tilde{H}_{-n} (-\mathbf{k}) = H_{eff} (-\mathbf{k},\theta) .
               \label{C6}
            \end{equation}
          \par

          For PHS,
            \begin{equation}
               \mathcal{C} \mathbf{c}_{\mathbf{k}} \mathcal{C}^{-1} = U_{\mathcal{C}}^{*} \mathbf{c}_{-\mathbf{k}}^{\dagger}  ,
               \qquad
               \mathcal{C} i \mathcal{C}^{-1} = i ,
               \label{C7}
            \end{equation}
          where, $U_{\mathcal{C}}$ is a unitary matrix. Thus,
            \begin{equation}
               \begin{split}
                  H_{\mathcal{C}} & = \mathcal{C} H \mathcal{C}^{-1} = \sum_{\mathbf{k}} \sum_{n=-\infty}^{\infty} e^{-i n \Omega t} \mathcal{C} \tilde{H}_n (\mathbf{k}) \mathcal{C}^{-1} \\
                   & = \sum_{\mathbf{k}} \sum_{n=-\infty}^{\infty} e^{-i n \Omega t} \sum_{ \tilde{\mathbf{R}}} e^{-i \mathbf{k} \cdot \tilde{\mathbf{R}}} \mathbf{c}_{-\mathbf{k}} U_{\mathcal{C}}^{T} \mathbf{M}_{\tilde{\mathbf{R}}}^{(n)} U_{\mathcal{C}}^{*} \mathbf{c}_{-\mathbf{k}}^{\dagger} \\
                    =& - \sum_{\mathbf{k}} \sum_{n=-\infty}^{\infty} e^{-i n \Omega t} \sum_{ \tilde{\mathbf{R}}} e^{-i \mathbf{k} \cdot \tilde{\mathbf{R}}} \mathbf{c}_{-\mathbf{k}}^{\dagger} U_{\mathcal{C}}^{\dagger} \mathbf{M}_{\tilde{\mathbf{R}}}^{(n)T} U_{\mathcal{C}} \mathbf{c}_{-\mathbf{k}}  \\
                    & = \sum_{\mathbf{k}} \sum_{n=-\infty}^{\infty} e^{-i n \Omega t} \tilde{H}_{n,\mathcal{C}} (- \mathbf{k}) ,
               \end{split}
               \label{C8}
            \end{equation}
          where 
            \begin{equation}
               \tilde{H}_{n,\mathcal{C}} (- \mathbf{k}) = - \sum_{ \tilde{\mathbf{R}}} e^{-i \mathbf{k} \cdot \tilde{\mathbf{R}}} \mathbf{c}_{-\mathbf{k}}^{\dagger} U_{\mathcal{C}}^{\dagger} \mathbf{M}_{\tilde{\mathbf{R}}}^{(n)T} U_{\mathcal{C}} \mathbf{c}_{-\mathbf{k}} ,
               \label{C9}
            \end{equation}
          and we assume that $TrH = 0$. The effective Hamiltonian corresponding to $H_{\mathcal{C}}$ is 
            \begin{equation}
               \mathcal{C} H_{eff} (\mathbf{k},\theta) \mathcal{C}^{-1} = H_{eff,\mathcal{C}} (\mathbf{k},\theta) = \sum_{n=-\infty}^{\infty} e^{- i n \theta}  \tilde{H}_{-n,\mathcal{C}} (\mathbf{k}).
               \label{C10}
            \end{equation}
          If the system has PHS, $H =- H_{\mathcal{C}}$. Thus, $\tilde{H}_{n,\mathcal{C}} (-\mathbf{k}) = - \tilde{H}_{n}(\mathbf{k})$ and 
            \begin{equation}
               \mathcal{C} H_{eff} (\mathbf{k},\theta) \mathcal{C}^{-1} = -\sum_{n=-\infty}^{\infty} e^{- i n \theta} \tilde{H}_{-n} (\mathbf{-k}) = -H_{eff} (-\mathbf{k},\theta).
               \label{C11}
            \end{equation}
          \par

          Since CS is the combination of TRS and PHS,
            \begin{equation}
               \mathcal{S} H_{eff} (\mathbf{k},\theta) \mathcal{S}^{-1} = -H_{eff} (\mathbf{k},\theta).
               \label{C12}
            \end{equation}

         \subsection{For Non-Hermitian case}
           We consider TRS firstly. If a non-Hermitian system has TRS,
             \begin{equation}
               \mathcal{T}_{+} H \mathcal{T}_{+}^{-1} = H,
               \label{C13}
             \end{equation}
           with 
             \begin{equation}
               \mathcal{T}_{+} \mathbf{c}_{\mathbf{k}} \mathcal{T}_{+}^{-1} = U_{\mathcal{T}_{+}} \mathbf{c}_{-\mathbf{k}}  ,
              \qquad
              \mathcal{T}_{+} i \mathcal{T}_{+}^{-1} = -i,
              \label{C14}
             \end{equation}
           and $ U_{\mathcal{T}_{+}}$ is a unitary matrix. TRS in non-Hermitian system has the same form as TRS in Hermitian system.  Thus, for non-Hermitian system with TRS,
             \begin{equation}
               \mathcal{T}_{+} H_{eff} (\mathbf{k},\theta) \mathcal{T}_{+}^{-1} = H_{eff,\mathcal{T}_{+}} (\mathbf{k},\theta) = H_{eff} (-\mathbf{k},\theta).
               \label{C15}
             \end{equation}
           For non-Hermitian PHS,
             \begin{equation}
               \mathcal{C}_{-} \mathbf{c}_{\mathbf{k}} \mathcal{C}_{-}^{-1} = U_{\mathcal{C}_{-}}^{*} \mathbf{c}_{-\mathbf{k}}^{\dagger} , 
               \qquad
               \mathcal{C}_{-} i \mathcal{C}_{-}^{-1} = i ,
               \label{C16}
             \end{equation}
           with a unitary matrix $U_{\mathcal{C}_{-}}$, and
              \begin{equation}
               \mathcal{C}_{-} H \mathcal{C}_{-}^{-1} = - H.
               \label{C17}
              \end{equation}
           The form of PHS in Hermitian system and non-Hermitian system are the same. Hence, for non-Hermitian PHS,
             \begin{equation}
               \mathcal{C}_{-} H_{eff} (\mathbf{k},\theta) \mathcal{C}_{-}^{-1} = H_{eff,\mathcal{C}_{-}} (\mathbf{k},\theta) = -H_{eff} (-\mathbf{k},\theta).
             \end{equation}
           \par

           Now, we consider TRS$^{\dagger}$ in non-Hermitian system. Under TRS$^{\dagger}$
             \begin{equation}
               \mathcal{C}_{+} \mathbf{c}_{\mathbf{k}} \mathcal{C}_{+}^{-1} = U_{\mathcal{C}_{+}} \mathbf{c}_{-\mathbf{k}}  ,
              \qquad
              \mathcal{C}_{+} i \mathcal{C}_{+}^{-1} = -i,
              \label{C19}
             \end{equation}
           with a unitary matrix $U_{\mathcal{C}_{+}}$. However, if a non-Hermitian system has TRS$^{\dagger}$, the Hamiltonian has the relation
             \begin{equation}
               \mathcal{C}_{+} H^{\dagger} \mathcal{C}_{+}^{-1} = H.
               \label{C20}
             \end{equation}
           For the Hamiltonian $H$ given in Eq.\eqref{C1} ,
             \begin{equation}
                \begin{split}
                  H_{\mathcal{C}_{+}} & = \mathcal{C}_{+} H \mathcal{C}_{+}^{-1} = \sum_{\mathbf{k}} \sum_{n=-\infty}^{\infty} e^{i n \Omega t} \mathcal{C}_{+} \tilde{H}_n (\mathbf{k}) \mathcal{C}_{+}^{-1}   \\
                  & =  \sum_{\mathbf{k}} \sum_{n=-\infty}^{\infty} e^{i n \Omega t} \sum_{ \tilde{\mathbf{R}}} e^{i \mathbf{k} \cdot \tilde{\mathbf{R}}} \mathbf{c}_{-\mathbf{k}}^{\dagger} U_{\mathcal{C}_{+}}^{\dagger} \mathbf{M}_{\tilde{\mathbf{R}}}^{(n)*} U_{\mathcal{C}_{+}} \mathbf{c}_{-\mathbf{k}}  \\
                  & = \sum_{\mathbf{k}} \sum_{n=-\infty}^{\infty} e^{- i n \Omega t} \tilde{H}_{n,\mathcal{C}_{+}} (-\mathbf{k}),
                \end{split}
                \label{C21}
             \end{equation}
           where 
             \begin{equation}
               \tilde{H}_{n,\mathcal{C}_{+}} (-\mathbf{k}) = \sum_{ \tilde{\mathbf{R}}} e^{i \mathbf{k} \cdot \tilde{\mathbf{R}}} \mathbf{c}_{-\mathbf{k}}^{\dagger} U_{\mathcal{C}_{+}}^{\dagger} \mathbf{M}_{\tilde{\mathbf{R}}}^{(-n)*} U_{\mathcal{C}_{+}} \mathbf{c}_{-\mathbf{k}}.
               \label{C22}
             \end{equation}
           Thus, the effective Hamiltonian corresponding to $H_{\mathcal{C}_{+}}$ is 
             \begin{equation}
               \mathcal{C}_{+} H_{eff} (\mathbf{k},\theta) \mathcal{C}_{+}^{-1} = H_{eff,\mathcal{C}_{+}} (\mathbf{k},\theta) = \sum_{n=-\infty}^{\infty} e^{- i n \theta}  \tilde{H}_{-n,\mathcal{C}_{+}} (\mathbf{k}).
               \label{C23}
             \end{equation}
           According to Eqs.\eqref{C1},\eqref{C20} and \eqref{C21}, $\tilde{H}_{n,\mathcal{C}_{+}} (-\mathbf{k}) = \tilde{H}_{-n}^{\dagger}(\mathbf{k})$. Hence,
             \begin{equation}
               \mathcal{C}_{+} H_{eff} (\mathbf{k},\theta) \mathcal{C}_{+}^{-1} = \sum_{n=-\infty}^{\infty} e^{- i n \theta} \tilde{H}_{n}^{\dagger}(-\mathbf{k}) = H_{eff}^{\dagger} (-\mathbf{k},\theta) .
               \label{C24}
             \end{equation}
           If a non-Hermitian system has PHS$^{\dagger}$,
             \begin{equation}
               \mathcal{T}_{-} H^{\dagger} \mathcal{T}_{-}^{-1} = -H ,
               \label{C25}
             \end{equation} 
           for the Hamiltonian $H$ and 
             \begin{equation}
               \mathcal{T}_{-} \mathbf{c}_{\mathbf{k}} \mathcal{T}_{-}^{-1} = U_{\mathcal{T}_{-}}^{*} \mathbf{c}_{-\mathbf{k}}^{\dagger}  ,
               \qquad
               \mathcal{T}_{-} i \mathcal{T}_{-}^{-1} = i ,
               \label{C26}
             \end{equation}
           with a unitary matrix $U_{\mathcal{T}_{-}}$. For $H$ given in Eq.\eqref{C1},
             \begin{equation}
                \begin{split}
                  H_{\mathcal{T}_{-}} & = \mathcal{T}_{-} H \mathcal{T}_{-}^{-1} = \sum_{\mathbf{k}} \sum_{n=-\infty}^{\infty} e^{-i n \Omega t} \mathcal{T}_{-} \tilde{H}_n (\mathbf{k}) \mathcal{T}_{-}^{-1}   \\
                   & = \sum_{\mathbf{k}} \sum_{n=-\infty}^{\infty} e^{-i n \Omega t} \sum_{ \tilde{\mathbf{R}}} e^{-i \mathbf{k} \cdot \tilde{\mathbf{R}}} \mathbf{c}_{-\mathbf{k}} U_{\mathcal{T}_{-}}^{T} \mathbf{M}_{\tilde{\mathbf{R}}}^{(n)} U_{\mathcal{T}_{-}}^{*} \mathbf{c}_{-\mathbf{k}}^{\dagger}  \\
                   & = - \sum_{\mathbf{k}} \sum_{n=-\infty}^{\infty} e^{-i n \Omega t} \sum_{ \tilde{\mathbf{R}}} e^{-i \mathbf{k} \cdot \tilde{\mathbf{R}}} \mathbf{c}_{-\mathbf{k}}^{\dagger} U_{\mathcal{T}_{-}}^{\dagger} \mathbf{M}_{\tilde{\mathbf{R}}}^{(n)T} U_{\mathcal{T}_{-}} \mathbf{c}_{-\mathbf{k}} \\
                   & = \sum_{\mathbf{k}} \sum_{n=-\infty}^{\infty} e^{-i n \Omega t} \tilde{H}_{n,\mathcal{T}_{-}} (- \mathbf{k}) ,
                \end{split}
                \label{C27}
             \end{equation}
           where
             \begin{equation}
               \tilde{H}_{n,\mathcal{T}_{-}} (- \mathbf{k}) = - \sum_{ \tilde{\mathbf{R}}} e^{-i \mathbf{k} \cdot \tilde{\mathbf{R}}} \mathbf{c}_{-\mathbf{k}}^{\dagger} U_{\mathcal{T}_{-}}^{\dagger} \mathbf{M}_{\tilde{\mathbf{R}}}^{(n)T} U_{\mathcal{T}_{-}} \mathbf{c}_{-\mathbf{k}}.
               \label{C28}
             \end{equation}
            The effective Hamiltonian corresponding to $\tilde{H}_{n,\mathcal{T}_{-}}$ is 
              \begin{equation}
               \mathcal{T}_{-} H_{eff} (\mathbf{k},\theta) \mathcal{T}_{-}^{-1} = H_{eff,\mathcal{T}_{-}} (\mathbf{k},\theta) = \sum_{n=-\infty}^{\infty} e^{- i n \theta} \tilde{H}_{-n,\mathcal{T}_{-}} ( \mathbf{k}).
               \label{C29}
              \end{equation}
           Since $H_{\mathcal{T}_{-}} = - H^{\dagger}$, $\tilde{H}_{n,\mathcal{T}_{-}} (- \mathbf{k}) =- H_{-n}^{\dagger} (\mathbf{k})$. Hence,
              \begin{equation}
               H_{eff,\mathcal{T}_{-}} (\mathbf{k},\theta) = - \sum_{n=-\infty}^{\infty} e^{- i n \theta} H_{n}^{\dagger} (-\mathbf{k}) = - H_{eff}^{\dagger} (- \mathbf{k},\theta).
               \label{C30}
              \end{equation}
           \par

           Since CS is the combination of TRS and PHS, or the combination of TRS$^{\dagger}$ and PHS$^{\dagger}$,
             \begin{equation}
               \mathcal{S} H_{eff} (\mathbf{k},\theta) \mathcal{S}^{-1} = -H_{eff} (\mathbf{k},\theta).
               \label{C31}
             \end{equation}
           \par
         
           SLS is the generalization of CS in non-Hermitian system. For SLS,
             \begin{equation}
               \Gamma \mathbf{c}_{\mathbf{k}} \Gamma^{-1} = U_{\Gamma}^{*} \mathbf{c}_{\mathbf{k}}^{\dagger}  ,
               \qquad
               \Gamma i \Gamma^{-1} = -i,
               \label{C32}
             \end{equation}
           where $U_{\Gamma}$ is a unitary matrix with $U_{\Gamma}^2=\mathbf{1}$. If $H$ satisfies SLS,
             \begin{equation}
               \Gamma H \Gamma^{-1}  = H_{\Gamma} = -H^{\dagger}.
               \label{C33}
             \end{equation}
           According to Eq.\eqref{C32},
             \begin{equation}
                \begin{split}
                  H_{\Gamma} & = \Gamma H \Gamma^{-1} = \sum_{\mathbf{k}} \sum_{n=-\infty}^{\infty} e^{i n \Omega t} \Gamma \tilde{H}_n (\mathbf{k}) \Gamma^{-1}   \\
                  & = \sum_{\mathbf{k}} \sum_{n=-\infty}^{\infty} e^{i n \Omega t} \sum_{ \tilde{\mathbf{R}}} e^{i \mathbf{k} \cdot \tilde{\mathbf{R}}} \mathbf{c}_{\mathbf{k}} U_{\Gamma}^{T} \mathbf{M}_{\tilde{\mathbf{R}}}^{(n)*} U_{\Gamma}^{*} \mathbf{c}_{\mathbf{k}}^{\dagger}  \\
                  & = - \sum_{\mathbf{k}} \sum_{n=-\infty}^{\infty} e^{i n \Omega t} \sum_{ \tilde{\mathbf{R}}} e^{i \mathbf{k} \cdot \tilde{\mathbf{R}}} \mathbf{c}_{\mathbf{k}}^{\dagger} U_{\Gamma}^{\dagger} \mathbf{M}_{\tilde{\mathbf{R}}}^{(n)\dagger} U_{\Gamma} \mathbf{c}_{\mathbf{k}}  \\
                  & = \sum_{\mathbf{k}} \sum_{n=-\infty}^{\infty} e^{-i n \Omega t} \tilde{H}_{n,\Gamma} ( \mathbf{k}),
                \end{split}
                \label{C34}
             \end{equation}
           where
             \begin{equation}
              \tilde{H}_{n,\Gamma} ( \mathbf{k}) = \sum_{ \tilde{\mathbf{R}}} e^{i \mathbf{k} \cdot \tilde{\mathbf{R}}} \mathbf{c}_{\mathbf{k}}^{\dagger} U_{\Gamma}^{\dagger} \mathbf{M}_{\tilde{\mathbf{R}}}^{(-n)\dagger} U_{\Gamma} \mathbf{c}_{\mathbf{k}}.
               \label{C35}
             \end{equation}
           Then,
             \begin{equation}
               \Gamma H_{eff} (\mathbf{k},\theta) \Gamma^{-1} = H_{eff,\Gamma} (\mathbf{k},\theta) = \sum_{n=-\infty}^{\infty} e^{-i n \theta} \tilde{H}_{-n,\Gamma} (\mathbf{k}).
               \label{C36}
             \end{equation} 
            If $H$ satisfies Eq.\eqref{C33}, $\tilde{H}_{n,\Gamma} ( \mathbf{k}) = - \tilde{H}_{-n}^{\dagger} (\mathbf{k})$. Hence,
              \begin{equation}
               \Gamma H_{eff} (\mathbf{k},\theta) \Gamma^{-1} = - \sum_{n=-\infty}^{\infty} e^{-i n \theta} \tilde{H}_{n}^{\dagger} (\mathbf{k}) = - H_{eff}^{\dagger} (\mathbf{k},\theta).
               \label{C37}
              \end{equation}

           \par

           For pseudo-Hermitian symmetry, the operator $\eta$ is a linear operator~\cite{9}. Thus,
             \begin{equation}
               \eta \mathbf{c}_{\mathbf{k}} \eta^{-1} = U_{\eta} \mathbf{c}_{\mathbf{k}}  ,
               \qquad
               \eta i \eta^{-1} = i,
               \label{C38}
             \end{equation}
           where $U_{\eta}$ is a unitary and Hermitian matrix. If $H$ is pseudo-Hermitian,
             \begin{equation}
               \eta H \eta^{-1} = H^{\dagger}.
               \label{C39}
             \end{equation}
           We know that
             \begin{equation}
                \begin{split}
                  H_{\eta} & = \eta H \eta^{-1} = \sum_{\mathbf{k}} \sum_{n=-\infty}^{\infty} e^{-i n \Omega t} \eta \tilde{H}_{n} (\mathbf{k})\eta^{-1}  \\
                   & =  \sum_{\mathbf{k}} \sum_{n=-\infty}^{\infty} e^{-i n \Omega t} \sum_{ \tilde{\mathbf{R}}} e^{-i \mathbf{k} \cdot \tilde{\mathbf{R}}} \mathbf{c}_{\mathbf{k}}^{\dagger} U_{\eta}^{\dagger} \mathbf{M}_{\tilde{\mathbf{R}}}^{(n)} U_{\eta} \mathbf{c}_{\mathbf{k}}  \\
                   & = \sum_{\mathbf{k}} \sum_{n=-\infty}^{\infty} e^{-i n \Omega t} \tilde{H}_{n,\eta} (\mathbf{k}),
                \end{split}
                \label{C40}
             \end{equation}
           where
             \begin{equation}
               \tilde{H}_{n,\eta} (\mathbf{k}) = \sum_{ \tilde{\mathbf{R}}} e^{-i \mathbf{k} \cdot \tilde{\mathbf{R}}} \mathbf{c}_{\mathbf{k}}^{\dagger} U_{\eta}^{\dagger} \mathbf{M}_{\tilde{\mathbf{R}}}^{(n)} U_{\eta} \mathbf{c}_{\mathbf{k}},
               \label{C41}
             \end{equation}
            and the effective Hamiltonian corresponding to $\tilde{H}_{n,\eta}$ is 
              \begin{equation}
               H_{eff,\eta} (\mathbf{k},\theta) = \eta H_{eff} (\mathbf{k},\theta) \eta^{-1} = \sum_{n=-\infty}^{\infty} e^{-i n \theta} \tilde{H}_{-n,\eta} (\mathbf{k})
              \end{equation}
           If $H$ satisfies Eq.\eqref{C39}, $\tilde{H}_{n,\eta} (\mathbf{k}) = \tilde{H}_{-n}^{\dagger} (\mathbf{k})$. Hence,
              \begin{equation}
               \eta H_{eff} (\mathbf{k},\theta) \eta^{-1} = \sum_{n=-\infty}^{\infty} e^{-i n \theta} \tilde{H}_{n}^{\dagger} (\mathbf{k}) = H_{eff}^{\dagger} (\mathbf{k},\theta).
              \end{equation}
            \par

            All above give the effective Hamiltonian transformed under internal symmetries.

      \section{The Constraints of Space-Time Translation Symmetry Given by TRS}
         Consider a $(d+1)$-dimensional lattice potential $V(\mathbf{r},t)$ with space-time translation symmetry, whose space-time translation vectors are $\{ (\mathbf{s}_i, \tau_i) \}$ for $i=1,2,\cdots, d+1$. That means
           \begin{equation}
              V(\mathbf{r},t) = V(\mathbf{r} + \mathbf{s}_i ,t + \tau_i) 
              \label{E1}
           \end{equation}
         for $i=1,2,\cdots, d+1$. If $V(\mathbf{r},t)$ is invariant under TRS, we have 
           \begin{equation}
              V(\mathbf{r},t) = \mathcal{T} V(\mathbf{r},t) = V^{*}(\mathbf{r},-t),
              \label{E2}
           \end{equation}
        where $\mathcal{T}$ is the time-reversal operator and ``$*$'' means complex conjugation. Since $V(\mathbf{r},t)$ satisfies Eq.\eqref{E1} and Eq.\eqref{E2} simultaneously,
          \begin{equation}
            V(\mathbf{r},t) = V^{*}(\mathbf{r},-t) = V^{*}(\mathbf{r} + \mathbf{s}_i ,-t + \tau_i) =V(\mathbf{r} + \mathbf{s}_i ,t - \tau_i) .
            \label{E3}
          \end{equation}
        Eq.\eqref{E3} shows that for a $(d+1)$-dimensional space-time crystal system with space-time translation vectors $\{ (\mathbf{s}_i, \tau_i) \}$~($i=1,2,\cdots, d+1$), if this system is invariant under TRS, $\{ (\mathbf{s}_i, -\tau_i) \}$~($i=1,2,\cdots, d+1$) are also space-time translation vectors. Under these constraints, TRS is compatible with space-time translation symmetry.

   \bibliography{spacetime}

\begin{thebibliography}{29}%
\makeatletter
\providecommand \@ifxundefined [1]{%
 \@ifx{#1\undefined}
}%
\providecommand \@ifnum [1]{%
 \ifnum #1\expandafter \@firstoftwo
 \else \expandafter \@secondoftwo
 \fi
}%
\providecommand \@ifx [1]{%
 \ifx #1\expandafter \@firstoftwo
 \else \expandafter \@secondoftwo
 \fi
}%
\providecommand \natexlab [1]{#1}%
\providecommand \enquote  [1]{``#1''}%
\providecommand \bibnamefont  [1]{#1}%
\providecommand \bibfnamefont [1]{#1}%
\providecommand \citenamefont [1]{#1}%
\providecommand \href@noop [0]{\@secondoftwo}%
\providecommand \href [0]{\begingroup \@sanitize@url \@href}%
\providecommand \@href[1]{\@@startlink{#1}\@@href}%
\providecommand \@@href[1]{\endgroup#1\@@endlink}%
\providecommand \@sanitize@url [0]{\catcode `\\12\catcode `\$12\catcode
  `\&12\catcode `\#12\catcode `\^12\catcode `\_12\catcode `\%12\relax}%
\providecommand \@@startlink[1]{}%
\providecommand \@@endlink[0]{}%
\providecommand \url  [0]{\begingroup\@sanitize@url \@url }%
\providecommand \@url [1]{\endgroup\@href {#1}{\urlprefix }}%
\providecommand \urlprefix  [0]{URL }%
\providecommand \Eprint [0]{\href }%
\providecommand \doibase [0]{https://doi.org/}%
\providecommand \selectlanguage [0]{\@gobble}%
\providecommand \bibinfo  [0]{\@secondoftwo}%
\providecommand \bibfield  [0]{\@secondoftwo}%
\providecommand \translation [1]{[#1]}%
\providecommand \BibitemOpen [0]{}%
\providecommand \bibitemStop [0]{}%
\providecommand \bibitemNoStop [0]{.\EOS\space}%
\providecommand \EOS [0]{\spacefactor3000\relax}%
\providecommand \BibitemShut  [1]{\csname bibitem#1\endcsname}%
\let\auto@bib@innerbib\@empty
%</preamble>
\bibitem [{\citenamefont {Chiu}\ \emph {et~al.}(2016)\citenamefont {Chiu},
  \citenamefont {Teo}, \citenamefont {Schnyder},\ and\ \citenamefont
  {Ryu}}]{10}%
  \BibitemOpen
  \bibfield  {author} {\bibinfo {author} {\bibfnamefont {C.-K.}\ \bibnamefont
  {Chiu}}, \bibinfo {author} {\bibfnamefont {J.~C.~Y.}\ \bibnamefont {Teo}},
  \bibinfo {author} {\bibfnamefont {A.~P.}\ \bibnamefont {Schnyder}},\ and\
  \bibinfo {author} {\bibfnamefont {S.}~\bibnamefont {Ryu}},\ }\bibfield
  {title} {\bibinfo {title} {Classification of topological quantum matter with
  symmetries},\ }\href {https://doi.org/10.1103/RevModPhys.88.035005}
  {\bibfield  {journal} {\bibinfo  {journal} {Rev. Mod. Phys.}\ }\textbf
  {\bibinfo {volume} {88}},\ \bibinfo {pages} {035005} (\bibinfo {year}
  {2016})}\BibitemShut {NoStop}%
\bibitem [{\citenamefont {Qi}\ and\ \citenamefont {Zhang}(2011)}]{11}%
  \BibitemOpen
  \bibfield  {author} {\bibinfo {author} {\bibfnamefont {X.-L.}\ \bibnamefont
  {Qi}}\ and\ \bibinfo {author} {\bibfnamefont {S.-C.}\ \bibnamefont {Zhang}},\
  }\bibfield  {title} {\bibinfo {title} {Topological insulators and
  superconductors},\ }\href {https://doi.org/10.1103/RevModPhys.83.1057}
  {\bibfield  {journal} {\bibinfo  {journal} {Rev. Mod. Phys.}\ }\textbf
  {\bibinfo {volume} {83}},\ \bibinfo {pages} {1057} (\bibinfo {year}
  {2011})}\BibitemShut {NoStop}%
\bibitem [{\citenamefont {Qi}\ \emph {et~al.}(2008)\citenamefont {Qi},
  \citenamefont {Hughes},\ and\ \citenamefont {Zhang}}]{12}%
  \BibitemOpen
  \bibfield  {author} {\bibinfo {author} {\bibfnamefont {X.-L.}\ \bibnamefont
  {Qi}}, \bibinfo {author} {\bibfnamefont {T.~L.}\ \bibnamefont {Hughes}},\
  and\ \bibinfo {author} {\bibfnamefont {S.-C.}\ \bibnamefont {Zhang}},\
  }\bibfield  {title} {\bibinfo {title} {Topological field theory of
  time-reversal invariant insulators},\ }\href
  {https://doi.org/10.1103/PhysRevB.78.195424} {\bibfield  {journal} {\bibinfo
  {journal} {Phys. Rev. B}\ }\textbf {\bibinfo {volume} {78}},\ \bibinfo
  {pages} {195424} (\bibinfo {year} {2008})}\BibitemShut {NoStop}%
\bibitem [{\citenamefont {Tuegel}\ \emph {et~al.}(2019)\citenamefont {Tuegel},
  \citenamefont {Chua},\ and\ \citenamefont {Hughes}}]{13}%
  \BibitemOpen
  \bibfield  {author} {\bibinfo {author} {\bibfnamefont {T.~I.}\ \bibnamefont
  {Tuegel}}, \bibinfo {author} {\bibfnamefont {V.}~\bibnamefont {Chua}},\ and\
  \bibinfo {author} {\bibfnamefont {T.~L.}\ \bibnamefont {Hughes}},\ }\bibfield
   {title} {\bibinfo {title} {Embedded topological insulators},\ }\href
  {https://doi.org/10.1103/PhysRevB.100.115126} {\bibfield  {journal} {\bibinfo
   {journal} {Phys. Rev. B}\ }\textbf {\bibinfo {volume} {100}},\ \bibinfo
  {pages} {115126} (\bibinfo {year} {2019})}\BibitemShut {NoStop}%
\bibitem [{\citenamefont {Po}\ \emph {et~al.}(2018)\citenamefont {Po},
  \citenamefont {Watanabe},\ and\ \citenamefont {Vishwanath}}]{14}%
  \BibitemOpen
  \bibfield  {author} {\bibinfo {author} {\bibfnamefont {H.~C.}\ \bibnamefont
  {Po}}, \bibinfo {author} {\bibfnamefont {H.}~\bibnamefont {Watanabe}},\ and\
  \bibinfo {author} {\bibfnamefont {A.}~\bibnamefont {Vishwanath}},\ }\bibfield
   {title} {\bibinfo {title} {Fragile topology and wannier obstructions},\
  }\href {https://doi.org/10.1103/PhysRevLett.121.126402} {\bibfield  {journal}
  {\bibinfo  {journal} {Phys. Rev. Lett.}\ }\textbf {\bibinfo {volume} {121}},\
  \bibinfo {pages} {126402} (\bibinfo {year} {2018})}\BibitemShut {NoStop}%
\bibitem [{\citenamefont {Alexandradinata}\ \emph {et~al.}(2011)\citenamefont
  {Alexandradinata}, \citenamefont {Hughes},\ and\ \citenamefont
  {Bernevig}}]{15}%
  \BibitemOpen
  \bibfield  {author} {\bibinfo {author} {\bibfnamefont {A.}~\bibnamefont
  {Alexandradinata}}, \bibinfo {author} {\bibfnamefont {T.~L.}\ \bibnamefont
  {Hughes}},\ and\ \bibinfo {author} {\bibfnamefont {B.~A.}\ \bibnamefont
  {Bernevig}},\ }\bibfield  {title} {\bibinfo {title} {Trace index and spectral
  flow in the entanglement spectrum of topological insulators},\ }\href
  {https://doi.org/10.1103/PhysRevB.84.195103} {\bibfield  {journal} {\bibinfo
  {journal} {Phys. Rev. B}\ }\textbf {\bibinfo {volume} {84}},\ \bibinfo
  {pages} {195103} (\bibinfo {year} {2011})}\BibitemShut {NoStop}%
\bibitem [{\citenamefont {Benalcazar}\ \emph {et~al.}(2017)\citenamefont
  {Benalcazar}, \citenamefont {Bernevig},\ and\ \citenamefont {Hughes}}]{16}%
  \BibitemOpen
  \bibfield  {author} {\bibinfo {author} {\bibfnamefont {W.~A.}\ \bibnamefont
  {Benalcazar}}, \bibinfo {author} {\bibfnamefont {B.~A.}\ \bibnamefont
  {Bernevig}},\ and\ \bibinfo {author} {\bibfnamefont {T.~L.}\ \bibnamefont
  {Hughes}},\ }\bibfield  {title} {\bibinfo {title} {Electric multipole
  moments, topological multipole moment pumping, and chiral hinge states in
  crystalline insulators},\ }\href {https://doi.org/10.1103/PhysRevB.96.245115}
  {\bibfield  {journal} {\bibinfo  {journal} {Phys. Rev. B}\ }\textbf {\bibinfo
  {volume} {96}},\ \bibinfo {pages} {245115} (\bibinfo {year}
  {2017})}\BibitemShut {NoStop}%
\bibitem [{\citenamefont {Benalcazar}\ \emph {et~al.}(2019)\citenamefont
  {Benalcazar}, \citenamefont {Li},\ and\ \citenamefont {Hughes}}]{17}%
  \BibitemOpen
  \bibfield  {author} {\bibinfo {author} {\bibfnamefont {W.~A.}\ \bibnamefont
  {Benalcazar}}, \bibinfo {author} {\bibfnamefont {T.}~\bibnamefont {Li}},\
  and\ \bibinfo {author} {\bibfnamefont {T.~L.}\ \bibnamefont {Hughes}},\
  }\bibfield  {title} {\bibinfo {title} {Quantization of fractional corner
  charge in ${C}_{n}$-symmetric higher-order topological crystalline
  insulators},\ }\href {https://doi.org/10.1103/PhysRevB.99.245151} {\bibfield
  {journal} {\bibinfo  {journal} {Phys. Rev. B}\ }\textbf {\bibinfo {volume}
  {99}},\ \bibinfo {pages} {245151} (\bibinfo {year} {2019})}\BibitemShut
  {NoStop}%
\bibitem [{\citenamefont {Liu}\ \emph {et~al.}(2019)\citenamefont {Liu},
  \citenamefont {Zhang}, \citenamefont {Ai}, \citenamefont {Gong},
  \citenamefont {Kawabata}, \citenamefont {Ueda},\ and\ \citenamefont
  {Nori}}]{18}%
  \BibitemOpen
  \bibfield  {author} {\bibinfo {author} {\bibfnamefont {T.}~\bibnamefont
  {Liu}}, \bibinfo {author} {\bibfnamefont {Y.-R.}\ \bibnamefont {Zhang}},
  \bibinfo {author} {\bibfnamefont {Q.}~\bibnamefont {Ai}}, \bibinfo {author}
  {\bibfnamefont {Z.}~\bibnamefont {Gong}}, \bibinfo {author} {\bibfnamefont
  {K.}~\bibnamefont {Kawabata}}, \bibinfo {author} {\bibfnamefont
  {M.}~\bibnamefont {Ueda}},\ and\ \bibinfo {author} {\bibfnamefont
  {F.}~\bibnamefont {Nori}},\ }\bibfield  {title} {\bibinfo {title}
  {Second-order topological phases in non-hermitian systems},\ }\href
  {https://doi.org/10.1103/PhysRevLett.122.076801} {\bibfield  {journal}
  {\bibinfo  {journal} {Phys. Rev. Lett.}\ }\textbf {\bibinfo {volume} {122}},\
  \bibinfo {pages} {076801} (\bibinfo {year} {2019})}\BibitemShut {NoStop}%
\bibitem [{\citenamefont {Kawabata}\ \emph {et~al.}(2021)\citenamefont
  {Kawabata}, \citenamefont {Shiozaki},\ and\ \citenamefont {Ryu}}]{19}%
  \BibitemOpen
  \bibfield  {author} {\bibinfo {author} {\bibfnamefont {K.}~\bibnamefont
  {Kawabata}}, \bibinfo {author} {\bibfnamefont {K.}~\bibnamefont {Shiozaki}},\
  and\ \bibinfo {author} {\bibfnamefont {S.}~\bibnamefont {Ryu}},\ }\bibfield
  {title} {\bibinfo {title} {Topological field theory of non-hermitian
  systems},\ }\href {https://doi.org/10.1103/PhysRevLett.126.216405} {\bibfield
   {journal} {\bibinfo  {journal} {Phys. Rev. Lett.}\ }\textbf {\bibinfo
  {volume} {126}},\ \bibinfo {pages} {216405} (\bibinfo {year}
  {2021})}\BibitemShut {NoStop}%
\bibitem [{\citenamefont {Fu}\ \emph {et~al.}(2021)\citenamefont {Fu},
  \citenamefont {Hu},\ and\ \citenamefont {Wan}}]{20}%
  \BibitemOpen
  \bibfield  {author} {\bibinfo {author} {\bibfnamefont {Y.}~\bibnamefont
  {Fu}}, \bibinfo {author} {\bibfnamefont {J.}~\bibnamefont {Hu}},\ and\
  \bibinfo {author} {\bibfnamefont {S.}~\bibnamefont {Wan}},\ }\bibfield
  {title} {\bibinfo {title} {Non-hermitian second-order skin and topological
  modes},\ }\href {https://doi.org/10.1103/PhysRevB.103.045420} {\bibfield
  {journal} {\bibinfo  {journal} {Phys. Rev. B}\ }\textbf {\bibinfo {volume}
  {103}},\ \bibinfo {pages} {045420} (\bibinfo {year} {2021})}\BibitemShut
  {NoStop}%
\bibitem [{\citenamefont {Okuma}\ \emph {et~al.}(2020)\citenamefont {Okuma},
  \citenamefont {Kawabata}, \citenamefont {Shiozaki},\ and\ \citenamefont
  {Sato}}]{21}%
  \BibitemOpen
  \bibfield  {author} {\bibinfo {author} {\bibfnamefont {N.}~\bibnamefont
  {Okuma}}, \bibinfo {author} {\bibfnamefont {K.}~\bibnamefont {Kawabata}},
  \bibinfo {author} {\bibfnamefont {K.}~\bibnamefont {Shiozaki}},\ and\
  \bibinfo {author} {\bibfnamefont {M.}~\bibnamefont {Sato}},\ }\bibfield
  {title} {\bibinfo {title} {Topological origin of non-hermitian skin
  effects},\ }\href {https://doi.org/10.1103/PhysRevLett.124.086801} {\bibfield
   {journal} {\bibinfo  {journal} {Phys. Rev. Lett.}\ }\textbf {\bibinfo
  {volume} {124}},\ \bibinfo {pages} {086801} (\bibinfo {year}
  {2020})}\BibitemShut {NoStop}%
\bibitem [{\citenamefont {Li}\ and\ \citenamefont {Wan}(2021)}]{22}%
  \BibitemOpen
  \bibfield  {author} {\bibinfo {author} {\bibfnamefont {H.}~\bibnamefont
  {Li}}\ and\ \bibinfo {author} {\bibfnamefont {S.}~\bibnamefont {Wan}},\
  }\bibfield  {title} {\bibinfo {title} {Homotopy invariant in time-reversal
  and twofold rotation symmetric systems},\ }\href
  {https://doi.org/10.1103/PhysRevB.104.045150} {\bibfield  {journal} {\bibinfo
   {journal} {Phys. Rev. B}\ }\textbf {\bibinfo {volume} {104}},\ \bibinfo
  {pages} {045150} (\bibinfo {year} {2021})}\BibitemShut {NoStop}%
\bibitem [{\citenamefont {Roy}\ and\ \citenamefont {Harper}(2017)}]{23}%
  \BibitemOpen
  \bibfield  {author} {\bibinfo {author} {\bibfnamefont {R.}~\bibnamefont
  {Roy}}\ and\ \bibinfo {author} {\bibfnamefont {F.}~\bibnamefont {Harper}},\
  }\bibfield  {title} {\bibinfo {title} {Periodic table for floquet topological
  insulators},\ }\href {https://doi.org/10.1103/PhysRevB.96.155118} {\bibfield
  {journal} {\bibinfo  {journal} {Phys. Rev. B}\ }\textbf {\bibinfo {volume}
  {96}},\ \bibinfo {pages} {155118} (\bibinfo {year} {2017})}\BibitemShut
  {NoStop}%
\bibitem [{\citenamefont {Fu}\ and\ \citenamefont {Kane}(2006)}]{25}%
  \BibitemOpen
  \bibfield  {author} {\bibinfo {author} {\bibfnamefont {L.}~\bibnamefont
  {Fu}}\ and\ \bibinfo {author} {\bibfnamefont {C.~L.}\ \bibnamefont {Kane}},\
  }\bibfield  {title} {\bibinfo {title} {Time reversal polarization and a
  ${Z}_{2}$ adiabatic spin pump},\ }\href
  {https://doi.org/10.1103/PhysRevB.74.195312} {\bibfield  {journal} {\bibinfo
  {journal} {Phys. Rev. B}\ }\textbf {\bibinfo {volume} {74}},\ \bibinfo
  {pages} {195312} (\bibinfo {year} {2006})}\BibitemShut {NoStop}%
\bibitem [{\citenamefont {Fu}\ and\ \citenamefont {Kane}(2007)}]{26}%
  \BibitemOpen
  \bibfield  {author} {\bibinfo {author} {\bibfnamefont {L.}~\bibnamefont
  {Fu}}\ and\ \bibinfo {author} {\bibfnamefont {C.~L.}\ \bibnamefont {Kane}},\
  }\bibfield  {title} {\bibinfo {title} {Topological insulators with inversion
  symmetry},\ }\href {https://doi.org/10.1103/PhysRevB.76.045302} {\bibfield
  {journal} {\bibinfo  {journal} {Phys. Rev. B}\ }\textbf {\bibinfo {volume}
  {76}},\ \bibinfo {pages} {045302} (\bibinfo {year} {2007})}\BibitemShut
  {NoStop}%
\bibitem [{\citenamefont {Borgnia}\ \emph {et~al.}(2020)\citenamefont
  {Borgnia}, \citenamefont {Kruchkov},\ and\ \citenamefont {Slager}}]{27}%
  \BibitemOpen
  \bibfield  {author} {\bibinfo {author} {\bibfnamefont {D.~S.}\ \bibnamefont
  {Borgnia}}, \bibinfo {author} {\bibfnamefont {A.~J.}\ \bibnamefont
  {Kruchkov}},\ and\ \bibinfo {author} {\bibfnamefont {R.-J.}\ \bibnamefont
  {Slager}},\ }\bibfield  {title} {\bibinfo {title} {Non-hermitian boundary
  modes and topology},\ }\href {https://doi.org/10.1103/PhysRevLett.124.056802}
  {\bibfield  {journal} {\bibinfo  {journal} {Phys. Rev. Lett.}\ }\textbf
  {\bibinfo {volume} {124}},\ \bibinfo {pages} {056802} (\bibinfo {year}
  {2020})}\BibitemShut {NoStop}%
\bibitem [{\citenamefont {Wu}\ and\ \citenamefont {An}(2020)}]{4}%
  \BibitemOpen
  \bibfield  {author} {\bibinfo {author} {\bibfnamefont {H.}~\bibnamefont
  {Wu}}\ and\ \bibinfo {author} {\bibfnamefont {J.-H.}\ \bibnamefont {An}},\
  }\bibfield  {title} {\bibinfo {title} {Floquet topological phases of
  non-hermitian systems},\ }\href {https://doi.org/10.1103/PhysRevB.102.041119}
  {\bibfield  {journal} {\bibinfo  {journal} {Phys. Rev. B}\ }\textbf {\bibinfo
  {volume} {102}},\ \bibinfo {pages} {041119} (\bibinfo {year}
  {2020})}\BibitemShut {NoStop}%
\bibitem [{\citenamefont {Yao}\ \emph {et~al.}(2017)\citenamefont {Yao},
  \citenamefont {Yan},\ and\ \citenamefont {Wang}}]{5}%
  \BibitemOpen
  \bibfield  {author} {\bibinfo {author} {\bibfnamefont {S.}~\bibnamefont
  {Yao}}, \bibinfo {author} {\bibfnamefont {Z.}~\bibnamefont {Yan}},\ and\
  \bibinfo {author} {\bibfnamefont {Z.}~\bibnamefont {Wang}},\ }\bibfield
  {title} {\bibinfo {title} {Topological invariants of floquet systems: General
  formulation, special properties, and floquet topological defects},\ }\href
  {https://doi.org/10.1103/PhysRevB.96.195303} {\bibfield  {journal} {\bibinfo
  {journal} {Phys. Rev. B}\ }\textbf {\bibinfo {volume} {96}},\ \bibinfo
  {pages} {195303} (\bibinfo {year} {2017})}\BibitemShut {NoStop}%
\bibitem [{\citenamefont {Shen}\ \emph {et~al.}(2018)\citenamefont {Shen},
  \citenamefont {Zhen},\ and\ \citenamefont {Fu}}]{6}%
  \BibitemOpen
  \bibfield  {author} {\bibinfo {author} {\bibfnamefont {H.}~\bibnamefont
  {Shen}}, \bibinfo {author} {\bibfnamefont {B.}~\bibnamefont {Zhen}},\ and\
  \bibinfo {author} {\bibfnamefont {L.}~\bibnamefont {Fu}},\ }\bibfield
  {title} {\bibinfo {title} {Topological band theory for non-hermitian
  hamiltonians},\ }\href {https://doi.org/10.1103/PhysRevLett.120.146402}
  {\bibfield  {journal} {\bibinfo  {journal} {Phys. Rev. Lett.}\ }\textbf
  {\bibinfo {volume} {120}},\ \bibinfo {pages} {146402} (\bibinfo {year}
  {2018})}\BibitemShut {NoStop}%
\bibitem [{\citenamefont {Yao}\ \emph {et~al.}(2018)\citenamefont {Yao},
  \citenamefont {Song},\ and\ \citenamefont {Wang}}]{7}%
  \BibitemOpen
  \bibfield  {author} {\bibinfo {author} {\bibfnamefont {S.}~\bibnamefont
  {Yao}}, \bibinfo {author} {\bibfnamefont {F.}~\bibnamefont {Song}},\ and\
  \bibinfo {author} {\bibfnamefont {Z.}~\bibnamefont {Wang}},\ }\bibfield
  {title} {\bibinfo {title} {Non-hermitian chern bands},\ }\href
  {https://doi.org/10.1103/PhysRevLett.121.136802} {\bibfield  {journal}
  {\bibinfo  {journal} {Phys. Rev. Lett.}\ }\textbf {\bibinfo {volume} {121}},\
  \bibinfo {pages} {136802} (\bibinfo {year} {2018})}\BibitemShut {NoStop}%
\bibitem [{\citenamefont {Kawabata}\ \emph {et~al.}(2019)\citenamefont
  {Kawabata}, \citenamefont {Shiozaki}, \citenamefont {Ueda},\ and\
  \citenamefont {Sato}}]{8}%
  \BibitemOpen
  \bibfield  {author} {\bibinfo {author} {\bibfnamefont {K.}~\bibnamefont
  {Kawabata}}, \bibinfo {author} {\bibfnamefont {K.}~\bibnamefont {Shiozaki}},
  \bibinfo {author} {\bibfnamefont {M.}~\bibnamefont {Ueda}},\ and\ \bibinfo
  {author} {\bibfnamefont {M.}~\bibnamefont {Sato}},\ }\bibfield  {title}
  {\bibinfo {title} {Symmetry and topology in non-hermitian physics},\ }\href
  {https://doi.org/10.1103/PhysRevX.9.041015} {\bibfield  {journal} {\bibinfo
  {journal} {Phys. Rev. X}\ }\textbf {\bibinfo {volume} {9}},\ \bibinfo {pages}
  {041015} (\bibinfo {year} {2019})}\BibitemShut {NoStop}%
\bibitem [{\citenamefont {Xu}\ and\ \citenamefont {Wu}(2018)}]{1}%
  \BibitemOpen
  \bibfield  {author} {\bibinfo {author} {\bibfnamefont {S.}~\bibnamefont
  {Xu}}\ and\ \bibinfo {author} {\bibfnamefont {C.}~\bibnamefont {Wu}},\
  }\bibfield  {title} {\bibinfo {title} {Space-time crystal and space-time
  group},\ }\href {https://doi.org/10.1103/PhysRevLett.120.096401} {\bibfield
  {journal} {\bibinfo  {journal} {Phys. Rev. Lett.}\ }\textbf {\bibinfo
  {volume} {120}},\ \bibinfo {pages} {096401} (\bibinfo {year}
  {2018})}\BibitemShut {NoStop}%
\bibitem [{\citenamefont {Peng}(2022)}]{2}%
  \BibitemOpen
  \bibfield  {author} {\bibinfo {author} {\bibfnamefont {Y.}~\bibnamefont
  {Peng}},\ }\bibfield  {title} {\bibinfo {title} {Topological space-time
  crystal},\ }\href {https://doi.org/10.1103/PhysRevLett.128.186802} {\bibfield
   {journal} {\bibinfo  {journal} {Phys. Rev. Lett.}\ }\textbf {\bibinfo
  {volume} {128}},\ \bibinfo {pages} {186802} (\bibinfo {year}
  {2022})}\BibitemShut {NoStop}%
\bibitem [{Note1()}]{Note1}%
  \BibitemOpen
  \bibinfo {note} {At least one number of $\{ m_i \}$ is nonzero.}\BibitemShut
  {Stop}%
\bibitem [{Note2()}]{Note2}%
  \BibitemOpen
  \bibinfo {note} {We take $\protect \mathbf {g}= diag \{ \protect \mathbf
  {1}_{d \times d} , -1 \}$ as the space-time metric to make it consistent with
  Refs.\cite {1},\cite {2}}\BibitemShut {NoStop}%
\bibitem [{\citenamefont {Sambe}(1973)}]{3}%
  \BibitemOpen
  \bibfield  {author} {\bibinfo {author} {\bibfnamefont {H.}~\bibnamefont
  {Sambe}},\ }\bibfield  {title} {\bibinfo {title} {Steady states and
  quasienergies of a quantum-mechanical system in an oscillating field},\
  }\href {https://doi.org/10.1103/PhysRevA.7.2203} {\bibfield  {journal}
  {\bibinfo  {journal} {Phys. Rev. A}\ }\textbf {\bibinfo {volume} {7}},\
  \bibinfo {pages} {2203} (\bibinfo {year} {1973})}\BibitemShut {NoStop}%
\bibitem [{\citenamefont {Chen}\ \emph {et~al.}(2018)\citenamefont {Chen},
  \citenamefont {Du},\ and\ \citenamefont {Fiete}}]{24}%
  \BibitemOpen
  \bibfield  {author} {\bibinfo {author} {\bibfnamefont {Q.}~\bibnamefont
  {Chen}}, \bibinfo {author} {\bibfnamefont {L.}~\bibnamefont {Du}},\ and\
  \bibinfo {author} {\bibfnamefont {G.~A.}\ \bibnamefont {Fiete}},\ }\bibfield
  {title} {\bibinfo {title} {Floquet band structure of a semi-dirac system},\
  }\href {https://doi.org/10.1103/PhysRevB.97.035422} {\bibfield  {journal}
  {\bibinfo  {journal} {Phys. Rev. B}\ }\textbf {\bibinfo {volume} {97}},\
  \bibinfo {pages} {035422} (\bibinfo {year} {2018})}\BibitemShut {NoStop}%
\bibitem [{\citenamefont {Mostafazadeh}(2002)}]{9}%
  \BibitemOpen
  \bibfield  {author} {\bibinfo {author} {\bibfnamefont {A.}~\bibnamefont
  {Mostafazadeh}},\ }\bibfield  {title} {\bibinfo {title} {Pseudo-hermiticity
  versus pt-symmetry iii: Equivalence of pseudo-hermiticity and the presence of
  antilinear symmetries},\ }\href {https://doi.org/10.1063/1.1489072}
  {\bibfield  {journal} {\bibinfo  {journal} {Journal of Mathematical Physics}\
  }\textbf {\bibinfo {volume} {43}},\ \bibinfo {pages} {3944} (\bibinfo {year}
  {2002})},\ \Eprint {https://arxiv.org/abs/https://doi.org/10.1063/1.1489072}
  {https://doi.org/10.1063/1.1489072} \BibitemShut {NoStop}%
\end{thebibliography}%


\end{document}